\documentclass[8pt,a4paper,usenatbib]{mnras}
\usepackage[utf8]{inputenc}
\usepackage{amsmath}
\usepackage{amsfonts}
\usepackage{amssymb}
\usepackage{graphicx}
\usepackage{lmodern}
\usepackage{multicol}
\usepackage{multirow}
\usepackage{hyperref}
\usepackage{comment}
\usepackage{verbatim}
\usepackage{pifont}
\usepackage{xspace}

\author[Pais \& Pfrommer]{Matteo Pais and Christoph Pfrommer\\
Leibniz-Institut f\"{u}r Astrophysik Potsdam,  An der Sternwarte 16, 14482 Potsdam, Germany \\
}
\title[Simulating TeV gamma-ray emission of SNRs]{Simulating TeV gamma-ray morphologies of shell-type supernova remnants}

\date{}
\newcommand{\de}{\mathrm{d}}

\newcommand{\erg}{\mathrm{erg}}

\renewcommand{\epsilon}{\varepsilon}

\newcommand{\AREPO}{\textsc{Arepo}\xspace}

\def\s{{\rm s}} %...........seconds
\def\yr{{\rm yr}} %.........years
\def\kyr{{\rm k}\yr} %......gigayears
\def\m{{\rm m}} %...........meters
\def\cm{{\rm c}\m} %........centimeters
\def\km{{\rm k}\m} %........kilometers
\def\pc{{\rm pc}} %.........parsecs
\def\kpc{{\rm k}\pc} %......kiloparsecs
\def\eV{{\rm eV}} %.........electron volts
\def\TeV{{\rm T}\eV} %......teraelectron volts
\def\erg{{\rm erg}} %.......ergs
\def\G{{\rm G}} %...........gauss
\def\muG{\mu\G} %...........microgauss
\def\mG{{\rm m}\G} %........milligauss
\def\bohm{{\rm Bohm}}  %...bohm
\def\prec{{\rm prec}}  %...precursor
\def\age{{\rm age}} %.....age
\def\tot{{\rm tot}}     %...total
\def\turb{{\rm turb}} %...turbulent
\usepackage[usenames]{color}
\definecolor{orange}{rgb}{1,0.5,0}

\newcommand{\vect}[1]{\boldsymbol{#1}}
\def\del#1{{}}

\begin{document}
\maketitle
\begin{abstract}
  Supernova remnant (SNR) shocks provide favourable sites of cosmic ray (CR)
  proton acceleration if the local magnetic field direction is quasi-parallel to
  the shock normal. Using the moving-mesh magneto-hydrodynamical (MHD) code
  \AREPO we present a suite of SNR simulations with CR acceleration in the
  Sedov-Taylor phase that combine different magnetic field topologies, density
  distributions with gradients and large-scale fluctuations, and -- for our
  core-collapse SNRs -- a multi-phase interstellar medium with dense clumps with
  a contrast of $10^4$. Assuming the hadronic gamma-ray emission model for the
  TeV gamma-ray emission, we find that large-amplitude density fluctuations of
  $\delta\rho/\rho_0\gtrsim75$ per cent are required to strongly modulate the
  gamma-ray emissivity in a straw man's model in which the acceleration
  efficiency is independent of magnetic obliquity. However, this causes strong
  corrugations of the shock surface that are ruled out by gamma-ray
  observations. By contrast, magnetic obliquity-dependent acceleration can
  easily explain the observed variance in gamma-ray morphologies ranging from
  SN1006 (with a homogeneous magnetic field) to Vela Junior and RX J1713 (with a
  turbulent field) in a single model that derives from plasma particle-in-cell
  simulations. Our best-fit model for SN1006 has a large-scale density gradient
  of $\nabla{n}\simeq0.0034~\cm^{-3}~\pc^{-1}$ pointing from south-west to
  north-east and a magnetic inclination with the plane of the sky of
  $\lesssim10^\circ$. Our best-fit model for Vela Junior and RX J1713 adopts a
  combination of turbulent magnetic field and dense clumps to explain their TeV
  gamma-ray morphologies and moderate shock corrugations.
\end{abstract}

\section{Introduction}
Ions are believed to be accelerated to relativistic energies at astrophysical
shocks, which gives rise to the observed CR population in the Galaxy
(\citealt{2005JPhG...31R..95H}, \citealt{2013A&ARv..21...70B} for a review). In
particular, the process of diffusive shock acceleration
\citep{1977ICRC...11..132A,1977DoSSR.234.1306K,1978ApJ...221L..29B,Bell1978a,
  Bell1978b} enables particles to gain energy through multiple shock crossings
as they scatter back and forth on magnetic field irregularities. The emerging
non-thermal spectrum follows a universal power-law momentum spectrum
\citep{Bell1978a,1978ApJ...221L..29B}. Supernova explosions and
subsequently formed remnant shocks are considered to be ideal environments for
acceleration because of the large spatial extent and lifetime that provides
sufficient confinement to reaching high energies \citep{2013SSRv..178..633G,
  Neronov2017}.  The most energetic CRs are able to escape upstream the shock
and propagate to larger distances in the interstellar medium (ISM) while
less energetic CRs are advected downstream and only released at a later time
\citep{2013MNRAS.431..415B}.

Evidence for efficient acceleration of CR electrons is provided by the
observation of elongated but narrow X-ray synchrotron filaments that are aligned
with the shock surface such as in Tycho
\citep{2002ApJ...581.1101H,2005ApJ...634..376W,2007ApJ...665..315C}, Vela
Jr.\ \citep{2005ApJ...632..294B}, or SN1006 \citep{2003ApJ...589..827B,
  2010ApJ...723..383K}, see also \citet{2006A&A...453..387P} for an overview.
Modelling the emission requires fast electron synchrotron losses in a strongly
amplified magnetic field in the upstream \citep[][for the case of
  SN1006]{2010MNRAS.405L..21M}, which is likely realised through the
non-resonant hybrid instability \citep{2004MNRAS.353..550B,2014ApJ...794...46C},
providing indirect evidence for efficient proton acceleration. This evidence is
further strengthened by multi-wavelength analyses that take into account the SNR
evolution, hydrodynamics of the shock (assuming spherical symmetry), magnetic
field amplification and the dynamical backreaction of CRs and self-generated
magnetic turbulence on the shock
\citep{2003A&A...412L..11B,2010ApJ...708..965Z,2012A&A...538A..81M}.  The total
density jump as measured from far upstream to the downstream exceeds the
canonical limit of four (for an ideal gas with adiabatic index 5/3) due to the
increased compressibility of the additional relativistic pressure of CRs
\citep{1983ApJ...272..765C, 2011ApJ...734...85C, 2014ApJ...783...91C,
  2017MNRAS.465.4500P}. This implies a smaller distance between the contact
discontinuity (CD) and the forward shock (FS) as observed in Tycho
\citep{2005ApJ...634..376W} as well as in SN1006 \citep{2008ApJ...680.1180C},
where this CD-FS distance shows a distinctive azimuthal variation such that it
is shorter in the polar cap regions, which show efficient amplification of the
magnetic field due to efficient CR proton acceleration (in part, the CD-FS
distance is reduced by the Rayleigh-Taylor instability of the CD).

Of particular interest is the very high energy (VHE) gamma-ray emission from
SNRs in the GeV and TeV regimes \citep{2005JPhG...31R..95H}, which directly
probes the CR component without the need to model the magnetic
field. Exceptional examples are shell-type SNRs, such as SN1006, Vela Jr.\ and
RX J1713-3948.5 (RX J1713 for short). Gamma-ray emission associated with these
objects can be produced in two different models \citep{2016EPJWC.12104001G,
  2016RPPh...79d6901M}. In the hadronic model $\pi^0$ mesons are produced in
inelastic CR-gas interactions and decay into pairs of gamma rays
\citep{2011JCAP...05..026C}. In the leptonic model the gamma-ray radiation
arises from a combination of inverse Compton scattering of the cosmic microwave
background (CMB) and starlight photons off of the accelerated CR electrons and
non-thermal bremsstrahlung. Following the CR electron spectrum in
three-dimensional MHD simulations of SNRs and modelling the multi-frequency
spectrum and emission maps from the radio to gamma-rays suggests that the GeV
gamma-ray regime has a significant leptonic contribution while the TeV range
is dominated by hadronic gamma rays \citep{Winner}.

The leptonic model naturally produces hard gamma-ray spectra while such a
spectrum can also be obtained in the hadronic model when considering a clumpy
ISM. Magnetic insulation of these dense clumps only allows high-energy CR
protons to penetrate into the dense regions, which implies a substantial
hardening of the proton spectrum in comparison to the acceleration spectrum in
the diffuse ISM \citep{2014MNRAS.445L..70G, 2019MNRAS.487.3199C}.  Combining
synchrotron and inverse Compton fluxes in the leptonic model produces
volume-filling magnetic field strengths of $\approx(10-35)\,\mu$G
\citep{2016EPJWC.12104001G,Winner}. These are only in agreement with mG-field
strengths inferred from X-ray synchrotron filaments when assuming a clumpy
medium, arguing for a detailed study of SNR emission maps. By contrast, many
previous studies focused on matching the multi-frequency spectra with
spherically symmetric models of the SNR evolution and neglecting the diversity
of morphological appearances of the observed shell-type SNRs.

In fact, the orientation of the upstream magnetic field plays an important role
in the acceleration process. Self consistent hybrid particle-in-cell (PIC)
simulations show that a quasi-parallel configuration is much more efficient in
accelerating CR protons in comparison to a quasi-perpendicular shock
geometry \citep{2014ApJ...783...91C,2015ApJ...798L..28C}. These PIC
  simulations also show that the normalisation of the non-thermal spectrum
  decreases with time \citep{2014ApJ...794...46C}, which keeps the energy in the
  non-thermal tail saturated despite the linear increase of the maximum CR
  particle energy with time \citep{2014ApJ...794...47C} so that the maximum
  acceleration efficiency of CR protons at quasi-parallel shocks is limited to
  $\approx15$ per cent \citep{2014ApJ...783...91C}. The idea that the resulting
emission could depend on the direction of the magnetic field in the X-ray band
has also been discussed \citep{2004A&A...425..121R}.  In our previous papers
\citep{2018MNRAS.478.5278P, 2020MNRAS.496.2448P, Winner}, we find that the
global topology of the magnetic field is fundamental in reproducing the
diversity of observed gamma-ray emission that ranges from a bi-lobed to a patchy
morphology. While an approximately homogeneous magnetic field produces a
bi-lobed gamma-ray emission, a turbulent field manifests itself in a patchy
emission characteristics. In this scenario, gamma-ray bright regions result from
quasi-parallel shocks which are known to efficiently accelerate CR protons, and
gamma-ray dark regions point to quasi-perpendicular shock configurations.  In
the case of an extremely small magnetic correlation length the emission
approaches an isotropic emission, albeit with a reduced effective acceleration
efficiency \citep{2018MNRAS.478.5278P}.

The interaction of a supernova explosion with a gas cloud has been explored in
various works \citep{1974ApJ...188..501C, 2015ARep...59..690K} and more recently
with highly-evolved individual explosions \citep{2019MNRAS.482.1602Z}.  In
particular, interstellar turbulence and its effect on CR acceleration
\citep{2004ARA&A..42..275S} has been the subject of various studies in the past
years (\citealt{2004ARA&A..42..211E} for a review). Simulations of single SNe
have been performed in an inhomogeneous ISM \citep{2015MNRAS.451.2757W} as well
as in the presence of a clumpy circumstellar medium
\citep{2014MNRAS.437..976O,2020MNRAS.496.2448P}. The clumpiness of the medium is
often assumed to be the sole source of irregularity of the emission morphology,
especially for strongly asymmetric distributions of heavy elements like Si and
Fe in the surroundings of SNRs as confirmed by XMM-Newton measurements
\citep{2015MNRAS.453.3953L}.

Following the spirit of this work, we simulate how an inhomogeneous medium
affects and regulates the TeV gamma-ray morphology of SNRs in the Sedov-Taylor
phase. We perform our simulations with various magnetic field configurations
combined with magnetic obliquity-dependent shock acceleration.  In order to
infer morphological properties of the ambient medium of observed SNRs, we
compare the resulting emission maps, radial and azimuthal profiles with three
well-known examples of shell-type SNRs such as SN1006, Vela Jr. and RX J1713.
For simplicity we use an initial setup of a point explosion that evolves into
the Sedov-Taylor solution. The gamma-ray emission from the resulting SNR is then
computed using a hadronic model of decaying pions. Our goal is to find a
consistent model that simultaneously explains the detailed gamma-ray spectrum
as well as the morphological variance of the TeV gamma-ray emission maps.

This paper is organised as follows. In Section~\ref{sec: methodology} we present
the methodology used to prepare our initial conditions for the various models
with particular focus on the generation of an initial turbulent density and
magnetic field. In Section~\ref{sec: Theoretical models} we present our suite of
SNR simulations with different combinations of magnetic fields (homogeneous and
turbulent) and density distributions (homogeneous, with a gradient, turbulent
and a combination of a gradient and turbulence), yielding a wide range of
different morphologies. For comparison, we also test a straw man's model of an
isotropic acceleration scenario (i.e., independent of magnetic obliquity). In
Section~\ref{sec: SN1006} we compare the observed gamma-ray map, radial and
azimuthal profiles of SN1006 with our simulations for different degrees of
turbulence and inclinations of the field of view and infer properties on the
local ISM. In Section~\ref{sec:CC_SNRs} we compare radial and azimuthal profiles
of turbulent SNRs Vela Jr. and RX J1713 to the models with a turbulent magnetic
field and a non-obliquity dependent acceleration for various density setups. In
Section~\ref{sec: high turbulence}, we study the effect of strong density
fluctuations on the gamma-ray morphology of a core-collapse SNR in its early
Sedov stage. In Section~\ref{sec: spectra}, we show that we can match the
gamma-ray spectra of all three SNRs for our adopted parameters and summarise the
main findings and conclude in Section~\ref{sec: conclusions}.

\section{Methodology}
\label{sec: methodology}
Here we present our methodology and briefly explain the procedure used to
implement obliquity dependent CR acceleration, the turbulence in the ISM and in
the initial magnetic field. Results of this setup are shown in Section~\ref{sec:
  Theoretical models}.

\subsection{Simulation method}
The simulations presented in this paper are performed with the massively
parallel, adaptive moving mesh-code \AREPO \citep{2010MNRAS.401..791S}. We use
an improved second-order hydrodynamic scheme with the least squares-fit gradient
estimate and a Runge-Kutta time integration \citep{Pakmor2016a}. Ideal MHD is
used to model the magnetic fields \citep{2013MNRAS.432..176P} while
zero-divergence is enforced through the implementation of a Powell scheme
\citep{1999JCoPh.154..284P}. CRs are modelled as a relativistic fluid with a
constant relativistic adiabatic index of 4/3 in a two-fluid approximation
\citep{2017MNRAS.465.4500P}.

Shocks are localised and characterised using the method developed by
\citet{2015MNRAS.446.3992S}, where Voronoi cells that exhibit a maximally
converging velocity field along the direction of propagation of the shock are
selected, while spurious shocks and numerical noise are filtered out. We inject
CR energy into the Voronoi cells in the immediate post-shock regime of shock
above a critical Mach number of $\mathcal{M}> 3$ \citep{2017MNRAS.465.4500P}.

Following the results of hybrid PIC simulations performed in
\citet{2014ApJ...783...91C} we assume a maximum CR energy efficiency of $15$ per
cent for quasi-parallel shocks. On the contrary quasi-perpendicular shocks are
found to be extremely inefficient accelerators. We note that the value of the
  maximum CR energy efficiency is uncertain. Arguments involving the Galactic CR
  energy budget to match the power needed to sustain the observed Galactic CR
  intensity favour average acceleration efficiencies of 3 to 10 per cent
  \citep{1989A&A...225..179D, 2006ApJ...636..777A, 2014arXiv1412.1376O,
    2015ASTRP...2...57K}, which can be scaled to maximum CR acceleration
  efficiencies at quasi-parallel shocks of 10 to 30 per cent
  \citep{2018MNRAS.478.5278P}, thus bracketing our value of 15 per cent.  The
efficiency of the accelerated CRs is computed using the orientation of the
pre-shock upstream magnetic field \citep{2018MNRAS.478.5278P}.

The presence of a strong current associated with streaming CRs into the upstream
region causes an exponential growth of the magnetic fluctuations via the
non-resonant hybrid instability \citep{2004MNRAS.353..550B}. These amplified
fluctuations saturate at wave amplitudes corresponding to the strength of the
mean magnetic field and cause the CR scattering mean free path to decrease to
values comparable with the fluctuating gyroradius. This regime approaches the
Bohm limit of diffusion. On scales resolved by our simulations, we neglect CR
diffusion and streaming. To justify this approach we assume Bohm diffusion and
calculate the CR precursor length for SNRs considered in this work and find
\begin{equation}
\begin{aligned}
L_{\prec} & \sim \sqrt{\kappa_\bohm t_\age} \simeq \\
& \simeq 0.3~\pc \left( \dfrac{\langle pc \rangle}{10~\TeV}\right)^{1/2}
\left( \dfrac{B}{10 \muG}  \right)^{-1/2} \left( \dfrac{t_\age}{10^3 \yr}\right)^{1/2} ,
\end{aligned}
\end{equation}
where $\langle pc \rangle$ is the average energy associated with the protons
that emit TeV gamma rays through the inelastic proton-proton reaction, $B$ is
the root-mean-square of the upstream magnetic field and $t_\age$ is the age of
the remnant. The resulting precursor length is below the size of our numerical
resolution $\Delta x = 0.4~\pc$, assuming a simulation box size of $40~\pc$ that
is filled with $100^3$ cells. Energy losses of accelerated CRs escaping upstream
from the blast wave do not affect the validity of the self-similar solution of
the problem, causing only a negligible softening of the Sedov-Taylor solution
\citep{2015MNRAS.447.2224B}. On the SNR timescales simulated, the hadronic and
Coulomb loss time scales are negligibly small so that we only account for
adiabatic CR losses.

\subsection{Initial conditions}
By analogy with \citet{2018MNRAS.478.5278P}, we start with a random Voronoi mesh
that is regularised into a glass-like configuration via Lloyd's algorithm
\citep{journals/tit/Lloyd82}.  We inject the equivalent of
$E_{\mathrm{SN}}=10^{51} \erg$ of thermal energy in the central cell of a
$100^3$-cell periodic box with a side length of $L=40~\pc$. The resulting
explosion forms an energy-driven strong shock expanding in an ISM characterised
by a low pressure of $0.44~\eV~\cm^{-3}$, and a mean molecular weight of
$\mu=1.4$.

\subsubsection{Magnetic field setup}
To generate a turbulent magnetic field we follow the procedure by
\citet{2017ApJ...844...13R} and \cite{2018MNRAS.478.5278P}. We adopt a
Kolmogorov-like power spectrum for the magnetic field and generate the three
magnetic vector components independently in Fourier space such that the
resulting field exhibits a random phase. We ensure that $\vect{B}$ is
divergence-free by projecting out the radial field component in Fourier
space. The degree of turbulence is determined through the fraction $f_B$ of
magnetic energy which goes into turbulent modes, yielding
\begin{equation}
  B_\tot^2 = B_0^2 +  B_\turb^2
\end{equation}
where $B_0^2 = B_\tot^2 ( 1- f_B)$ represents the strength of
  the mean field, $B_\turb^2 =\langle \delta \vect{b}^2 \rangle = f_B B_\tot^2 $
  is the average value of the Gaussian random field, and $\delta
  \vect{b}(\vect{x})$ is the local value of the turbulent field at the
  coordinates $\vect{x}=(x,y,z)$. Note that $0\leq f_B\leq1$ where $f_B =0$
  corresponds to a fully homogeneous field and $f_B = 1$ corresponds to a fully
  turbulent field.  We adopt a plasma beta factor of $\beta=1$ such that the
magnetic and thermal pressures are insignificant in comparison to the kinetic
energy of the propagating shock front. To maintain hydrostatic equilibrium in
the initial conditions, magnetic fluctuations $\delta \vect{B}(\vect{x})$ are
compensated by adopting temperature fluctuations of the form $n k_B \delta T
(\vect{x})= -\delta \vect{B}^2(\vect{x}) / (8 \pi)$. The small magnetic field
strength implies a small Alfv\'en speed so that the tension force is only slowly
mediated and does not affect the dynamics of our powerful shock wave.

\subsubsection{ISM density setup}
\label{sec:setup}
We model the multiphase structure of the ISM and adopt a combination of (i) a
large-scale linear density gradient, (ii) large-scale turbulent density
fluctuations that follow a Kolmogorov spectrum, and (iii) a population of small,
dense gaseous clumps with a typical overdensity of $10^4$ in comparison to the
ambient ISM.

The large-scale density gradient $\rho_\mathrm{grad}$ is determined by the
constant density $\rho_0$ and the slope parameter $\Delta$.  Similarly to the
magnetic field, we generate turbulent density fluctuations in Fourier space and
vary the initial seed, the correlation length and the amplitude of fluctuations
$\rho_\mathrm{turb}(\vect{x})=\rho_0 \delta \rho(\vect{x})$. To avoid negative
values for the density we adopt a density floor of $10^{-2} \rho_0$.  All these
elements are combined as follows:
\begin{align}
  \rho (\vect{x})  = &\rho_0 + \rho_\mathrm{turb}(\vect{x}) + \rho_\mathrm{grad}(\vect{x})
  = \rho_0 \left[1 + \delta \rho(\vect{x}) \right]\nonumber\\
  & +  \rho_2 \left[  \left(\frac{x}{L}-\frac{1}{2}\right) \cos\psi + \left(\frac{y}{L}-\frac{1}{2}\right) \sin\psi \right],
\label{eq: density}
\end{align}
where $L$ is the side length of our simulation box, $\psi$ represents the angle
between the direction of the density gradient and the $x-$axis,
$\delta\rho(\vect{x})$ is the generating functional of the turbulent field,
$\rho_2 = \sqrt{2} \Delta \rho_0 / (1+\Delta) $ is the amplitude of the gradient
such that $\rho = \rho_0 (1+ \rho_0 \delta \rho)$ for $\Delta= 0$ in case of a
vanishing gradient.  The function $\rho_\mathrm{grad}(\vect{x})$ is constructed
such that there are no negative values for the density in the simulation box for
any $x ,y \in [0,L]$ and $\Delta \in [0, \infty)$. To avoid low-density cavities
  in the surroundings of the central explosion, we shield the region with a
  constant homogeneous density of $\rho = \rho_0$ extending for a radius of
  $2~\pc$.

In order to reliably model the circum-stellar medium of our core-collapse SNRs,
we include a population of dense gaseous clumps with a typical size of 0.1~pc
and a number density of $\sim 10^{3} \cm^{-3}$
\citep{2012ApJ...744...71I}. Following the general setup as in
\citet{2020MNRAS.496.2448P} and \citet{2019MNRAS.487.3199C}, we include $7\times
10^3$ uniformly distributed small, dense clumps with a number density of
$n_\rmn{c} = 10^3~\cm^{-3}$ and a diameter of $0.1~\pc$ of a total target mass
of $M_\mathrm{c} = 45 M_\odot$ engulfed by the shock. Because of the
  quasi-Lagrangian property of our moving mesh code \AREPO, we resolve each
  small clump with initially 1000 cells that are uniformly distributed within
  the spherical clump volumes. As shown in Appendix~A of
  \citet{2020MNRAS.496.2448P}, our simulations accurately follow the dynamics of
  and inside every dense cloud that is site of vorticity generation upon the
  passage of the shock wave. As a result, Kelvin-Helmholtz instabilities start
  to mix cold, dense cloud material with the ambient hot shocked gas, which
  causes the formation of dense, ram-pressure stripped tails in the wake of each
  cloud \citep[see figure~2 of][]{2020MNRAS.496.2448P}.

This flexible setup allows us to study a wide range of different situations and
environments by modifying the amplitude of the density fluctuations, their
coherence scale and steepness of the gradient as well as their multi-phase
nature of thermally unstable dense clumps in the star-forming surroundings of
core-collapse SNRs.

\subsection{Modelling gamma-ray emission and noise}
To model the gamma-ray emission in post-processing we assume that the CR
population follows a universal power law momentum spectrum. A pion-decay
hadronic model is used to calculate the omni-directional gamma-ray emissivity
\citep{2004A&A...426..777P, 2008MNRAS.385.1211P} which depends on the local ISM
density, the local CR population, the very-high energy (i.e. $>1~\TeV$) photon
spectral index $\alpha_\gamma$, and the energy range. 

To compute the emission spectra of our simulations we integrate a particle
population described by a power law in momenta with cutoff. The hadronic
gamma-ray emission spectrum is calculated from parametrizations of the cross
section of neutral pion production at low and high proton energies, respectively
\citep{2018A&A...615A.108Y,2006PhRvD..74c4018K}.

In order to match the morphological properties of the observed gamma-ray
emission maps we proceed with an analysis of the power spectrum of the noise. We
fit the power spectrum with a specific function, convert it into a noise
map. This noise map is then superposed on the mock gamma-ray emission map that
was convolved with the observational point spread function (PSF). We will detail
the method to generate the noise properties in Section~\ref{sec: SN1006}.

\begin{figure*}
\centering\includegraphics[width=0.87\textwidth]{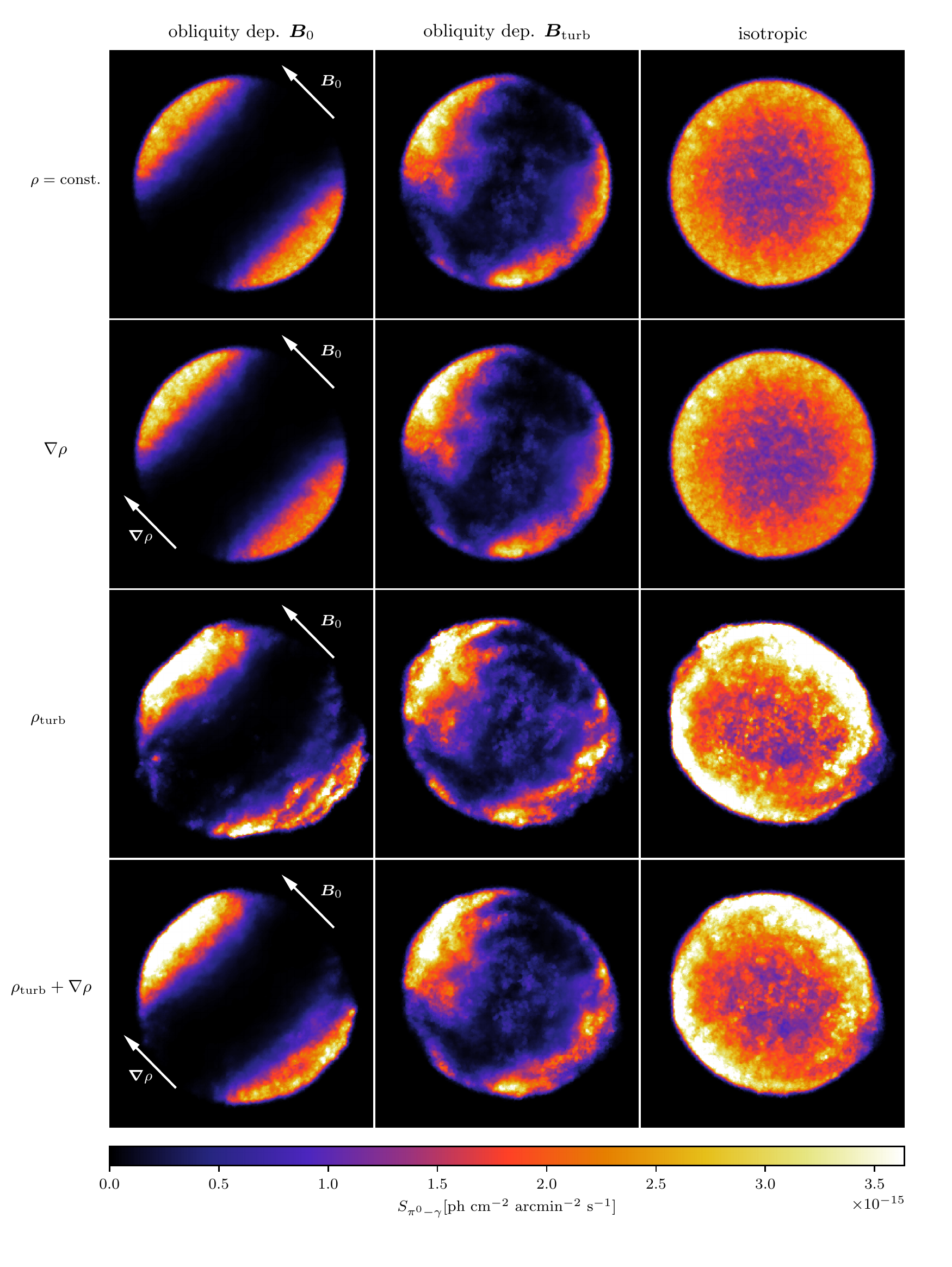} 
\caption{Impact of inhomogeneities density and magnetic field on the gamma-ray
  maps of SNRs in the Sedov-Taylor phase at $t_{\mathrm{age}} = 10^3 \yr$ with
  side length $L=40~\pc$. The images are ordered according to their different
  initial conditions: the magnetic field configurations and acceleration
    models differ between the three columns, and the adopted density
    distributions are different horizontally between each row. The first two
  columns assume magnetic obliquity dependent CR acceleration at a homogeneous
  magnetic field (left) and a fully turbulent fluctuations of relative amplitude
  $f_B=1$ and coherence length $\lambda_B=20~\pc$ (middle) while the third
  column shows simulations with a turbulent field ($\lambda_\rho=20~\pc$) but
  with an isotropic CR acceleration model. We adopt the following density
  distribution for all three simulations in each row: constant density
  (first row), an increasing gradient pointing from SW to NE (second row),
  turbulent density fluctuation (third row) and a combination of turbulence and
  a gradient (fourth row). The arrows indicate the direction of the magnetic
    field and the density gradient (if present).}
\label{Fig: 12_plots}
\end{figure*}

\section{Exploring fluctuations in density and the magnetic field}
\label{sec: Theoretical models}
Using the previously described setup here we present a suite of twelve SNR
simulations (each with $100^3$ cells) and combine different topologies of the
magnetic field and different density distributions. We also compare models of
obliquity-dependent CR acceleration to isotropic CR acceleration. The gamma-ray
emission maps in the energy range $1-80$~TeV with a spectral index of
$\alpha=2.1$ are shown in Fig.~\ref{Fig: 12_plots} at a SNR age of $t = 1000$
yrs.  The images are not convolved with a PSF and we do not add noise in order
to maintain the underlying setup as transparent as possible.

For these theoretical models we chose a simple setup both for the density and
the magnetic field. We start by separately considering the components of the
density distribution of Eqn.~\eqref{eq: density}.  For the background density we
select a value of $n=0.1~\cm^{-3}$. The density gradient has an inclination of
$\psi = 135^\circ$ with respect to the $x$-axis, i.e., it is pointing from
south-west (SW) to north-east (NE) and exhibits a moderate gradient intensity of
$\Delta = 1$ to avoid a strong contrast between the upper-left quadrant and the
bottom-right one. For the turbulence we chose turbulent fluctuations of $f_\rho
=\delta \rho / \rho_0 = 0.5$ and a coherence scale of $\lambda_\rho = L / 2$
for our box size $L=40~\pc$. To avoid any correlation between magnetic and
density fluctuations in our turbulent simulations, we used two distinct but
fixed random seeds, respectively.

\subsection{Density fluctuations}
In the first column of Fig.~\ref{Fig: 12_plots} we show the models for a
constant magnetic field oriented with an angle of $\theta=135^\circ$ with
respect to the positive $x$-axis. The effect of the obliquity-dependent shock
acceleration in presence of an ordered magnetic field is manifested in the
bi-lobed morphology of the gamma-ray emission. The presence of a moderate
gradient in the second and fourth row results in NE lobe brighter than the SW
lobe as expected in the presence of more ISM material on which the shock
impinges so that the freshly accelerated CR protons find more target gas in a
hadronic scenario.

\subsection{Magnetic turbulence}
In the central column of Fig.~\ref{Fig: 12_plots} we show the effect of a fully
turbulent magnetic field ($f_B=1$) that is superposed on different density
distributions. As expected, the obliquity-dependent shock acceleration allows
the creation of a patchy distribution of the brightness echoing the distribution
of the underlying locally accelerated CR population
\citep{2020MNRAS.496.2448P}. In particular, the combination with a turbulent
density distribution (Fig.~\ref{Fig: 12_plots}, third row, centre) is
responsible for a noticeable modulation of the gamma-ray intensity and explains
the bright spot in the lower right corner close to the remnant shell, an absent
feature in the constant density map (Fig.~\ref{Fig: 12_plots}, first row,
centre).

As expected in an isotropic acceleration scenario (Fig.~\ref{Fig: 12_plots},
right column) the gamma-ray emissivity shines along the entire SNR shell and is
only mildly modulated by local or large-scale variations related to the
particular density distribution used in the simulation. Contrarily to the
obliquity-dependent case for turbulent fields depicted in the central column, a
considerable amount of emission is still present in the centre of the remnant.

For isotropic models the emission in the centre is at least twice as strong as
the turbulent case. This is caused by the efficient CR acceleration in the
isotropic model so that the emission is projected onto the central parts of the
SNR. By contrast, in the obliquity-dependent scenario for the turbulent magnetic
case we obtain a patchy structured emission with bright spots in regions with a
quasi-parallel shock geometry and dark regions in locally quasi-parallel shocks
geometries. On average the acceleration process for medium to large scale
turbulence is $70$ per cent less efficient with respect to the non obliquity dependent
case \citep{2018MNRAS.478.5278P}. This results in a stronger dispersion of the
brightness and a lower average absolute emissivity in the centre. Note that
observational noise and the presence of a population of cold-dense clumps in a
multi-phase ISM may play an important role and needs to be taken in
consideration for the morphological modelling of observed shell-type SNRs.

\begin{table*}
\caption{Comparison of simulation and observational parameters for the presented hadronic morphological models.}
\begin{center}
\begin{tabular}{lcc|cc|cc}
\hline
& \multicolumn{2}{|c|}{\textbf{SN1006}} & \multicolumn{2}{|c|}{\textbf{Vela Jr.}} & \multicolumn{2}{|c|}{\textbf{RX J1713}} \\
\hline
\hline
\textbf{Parameter} & \textbf{Simulation} & \textbf{Observed} & \textbf{Simulation} & \textbf{Observed} & \textbf{Simulation} & \textbf{Observed}   \\ 
\hline
diameter $\theta_\mathrm{s}$ [deg] & 0.5 & 0.5 & 2 & 2 & 1 & 1 \\
$D$ [kpc] & 1.79  &  1.45 $-$ 2.2 & 0.5 & $0.2 - 0.75$ & 1 & $0.5 - 1$ \\
diameter $d_\mathrm{s}$ [pc] & 15.6 & 12.6 $-$19.2 & 17.4 & $7.0 - 26.1 $ & 17.4 & $8.7 - 17.4 $\\
$t_\age$ [kyr] & 1 & 1 & $2.7$ & $ 2.4 -5.1$ & $3$ & $1-7.9$ \\
$E_{\mathrm{SN}}~$[$10^{51}$ erg] & 1 & $-$ & 1 & $-$& 1 & $-$ \\
density $n~[\cm^{-3}]$ & 0.1 & 0.05 $-$ 0.3 & 0.42 & $0.03-1$ & $0.57$ & $0.02-1$ \\
$\nabla n~[\cm^{-3} \pc^{-1}]$ & 0.0034 &  $-$ & $-$ & $-$ & $0.02$ & $-$ \\
$ M_\mathrm{c}~[M_\odot]$ & $-$ & $-$ & 45 & $-$ & 45 & $-$ \\
 $\langle v_\mathrm{s} \rangle~[\km~\s^{-1}]$ & 3000 & $2100 - 4980$ & 2000 & $> 1000-3000$ & 1100 & $800-3900$ \\
$\mathcal{F}_\gamma(>1\mathrm{TeV})~[10^{-12}\mathrm{ph}~\mathrm{cm}^{-2}~\mathrm{s}^{-1}]$  & 0.4 & $0.39 \pm 0.08$ & 24 & $23.4 \pm 5.6 $ & 16.3 & $16.3 \pm 0.2$  \\
\hline
\hline
$\alpha_{\mathrm{p}, \mathrm{e}}$ & 1.95 & $ 1.79 \pm 0.44 $ &  1.81 & $1.85 \pm 0.24 $ & 1.7 & $1.52 \pm 0.31$ \\ 
$E_{\mathrm{p}, \mathrm{cut}} [\TeV]$ & 200 & $-$ & 100 & $-$& 90 & $-$ \\
$E_{\mathrm{e}, \mathrm{cut}} [\TeV]$ & $1.7$ & $-$& $0.25$ & $-$ & $0.25$ &\\
$\beta_\mathrm{p}$ & 2 & $-$ & 2 & $-$ & 2 & $-$ \\
$\beta_\mathrm{e}$ & 0.7 & $-$ & 0.4 & $-$ & 0.4 & $-$ \\
\hline
\hline
References & \multicolumn{2}{c}{1, 2, 6, 8, 9, 10, 13, 21} &  \multicolumn{2}{c}{2, 3, 4, 8, 10, 14, 16, 17, 19}  & \multicolumn{2}{c}{5, 7,  8, 11, 12, 15, 19, 20} \\
\hline
\end{tabular} \\[.5em]
\end{center}
Notes: $n$ denotes the diffuse ISM number density, $\nabla n$ is the large-scale density gradient,
$M_\mathrm{c}$ is the target clump mass hit by the remnant, $D$ is the distance to the SNR,
$\theta_{\rmn{s}}$ and $d_{\rmn{s}}$ are the angular and proper extent of the
blast wave, $\langle v_\mathrm{s} \rangle $ denotes the shock velocity, $t_\age$ is the SNR age and $\mathcal{F}_\gamma$ is the integrated gamma-ray flux above 1 TeV. 
References:  (1) \cite{2007AA...475..883A} ; (2) \cite{2015A&A...580A..74A}; (3) \cite{2015ApJ...798...82A}; (4) \cite{1999A&A...350..997A}; (5) \cite{2019MNRAS.487.3199C}; (6) \cite{2002A&A...387.1047D}; (7) \cite{2015A&A...577A..12F};  (8) \cite{2014BASI...42...47G}; (9) \cite{2010AA...516A..62A};  (10) \cite{2018AA...612A...7H}; (11) \cite{2018A&A...612A...6H}; (12) \cite{2012ApJ...744...71I}; (13) \cite{2008ApJ...678L..35K}; (14) \cite{2017hsn..book...63K}; (15) \cite{2020ApJS..248...16L}; (16) \cite{2019PhRvD.100b4063M}; (17) \cite{2001ApJ...548..814S}; (18) \cite{2008PASJ...60S.131T}
(19) \cite{2011ApJ...740L..51T}; (20) \cite{2016PASJ...68..108T}; (21) \cite{2003ApJ...585..324W}.
\label{table:Table 1}
\end{table*}

It is clear that the isotropic acceleration model fails to reproduce both the
emission morphologies (lobes and filaments) and the strong azimuthal variation
in the emission of SN1006. Despite a significantly varying amplitude of density
fluctuations of 50 per cent, the azimuthal variation in the isotropic-acceleration runs
does not drop sufficiently to create clearly defined filamentous structures in
the shell. High-amplitude turbulence in the ISM on scales comparable to the size
of the remnant might be a solution and could potentially mimic a bi-lobed
structure. We will come back to this point in Section~\ref{sec: high turbulence}
and will show that the almost spherically symmetric blast waves of the SNRs
studied here put a strong constraint on the level of large-scale inhomogeneity
of the surrounding ISM.

\begin{figure*}
\includegraphics[width=\textwidth]{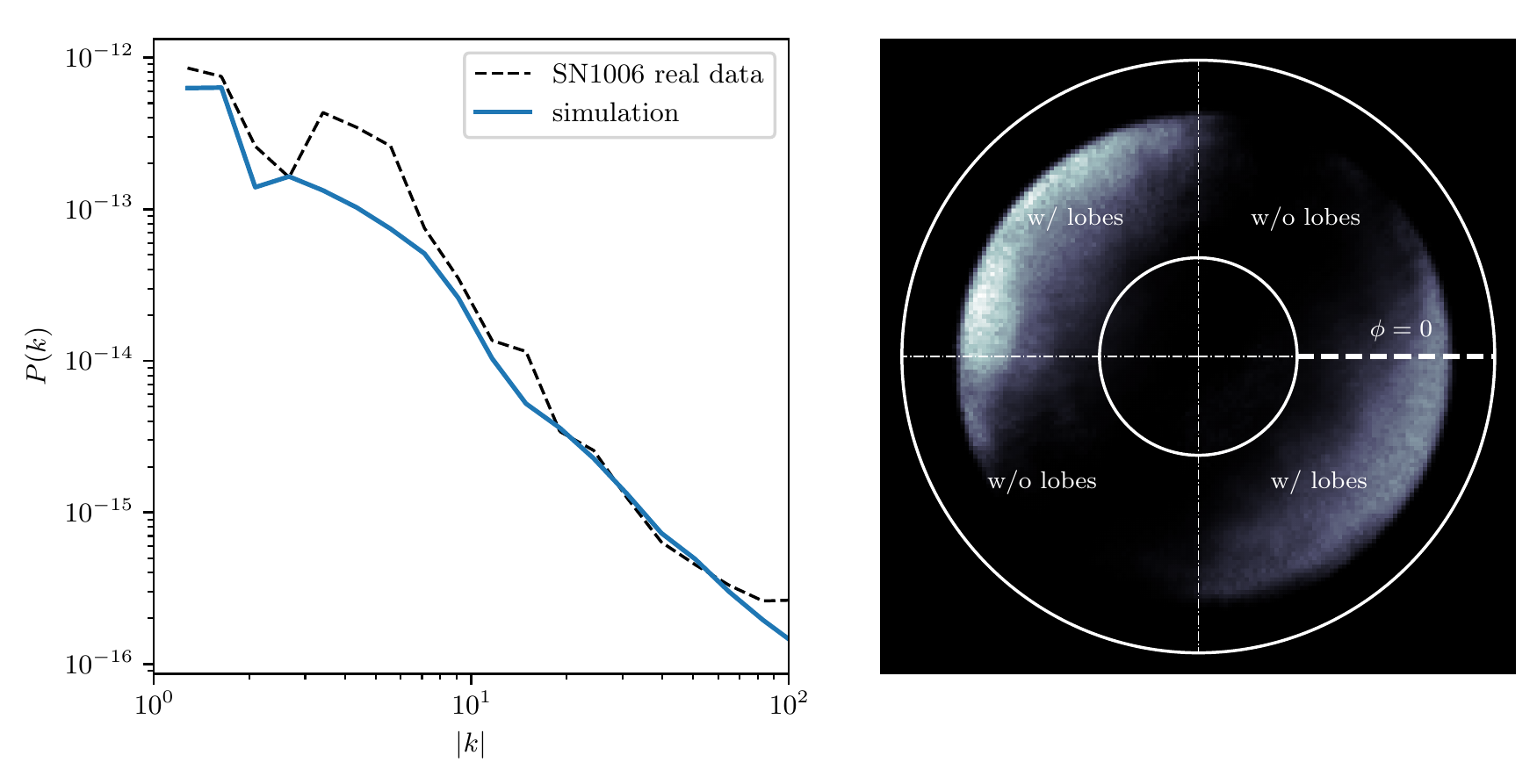} 
\caption{Left: Comparison of the noise power spectra of the observed emission map of
  SN1006 \citep{2010AA...516A..62A} with the signal regions masked and of the
  simulated gamma-ray map of SN1006 with a turbulent density fraction of
  $\delta\rho/\rho_0=0$. This shows that our modelled noise nicely
  corresponds to the observed noise properties. Right: Sketch of the area used
  for calculating the azimuthal and the radial profiles of the simulated
  models. We average the gamma-ray emission in the radial range $2/3 < r/\langle
  r_\rmn{s} \rangle < 4/3$ to obtain the azimuthal profile while the lobe
  regions are used to compute the radial profiles for SN1006.}
\label{Fig: power_spectra}
\end{figure*}

\begin{figure*}
\includegraphics[width=\textwidth]{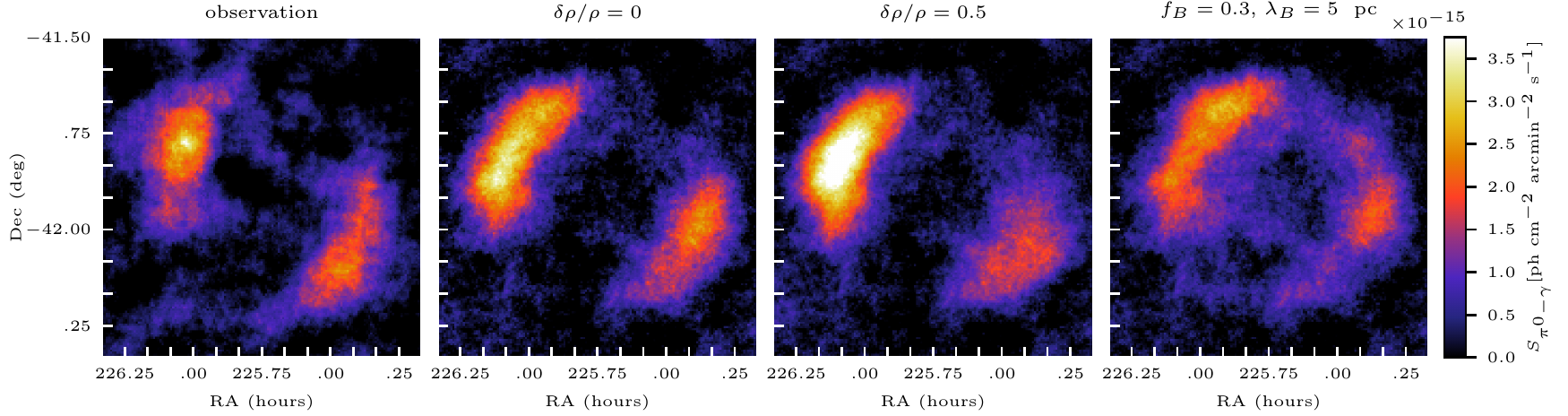} 
\caption{Two-dimensional projected gamma-ray maps for SNR SN1006 and for three
  simulation models with a fixed density gradient and different amplitudes of
  turbulent density fluctuations. The second and third panels show the simulated
  maps for $f_\rho = 5~\pc$ and $\delta\rho/\rho = 0 $ and $0.5$, respectively,
  while the rightmost panel shows the map for $f_B = 0.3$, $\lambda_B = 5~\pc$
  and $\delta \rho / \rho =0$. The emission in the simulated maps is smoothed
with a Gaussian PSF of width $\sigma=0.045^\circ$. On the top of it, we add
Gaussian noise with the power spectrum inferred from HESS data, as shown in
Fig.~\ref{Fig: power_spectra}, at fixed random seed.}
\label{Fig: SN1006_comparison}
\end{figure*}

\begin{figure*}
\includegraphics[width=\textwidth]{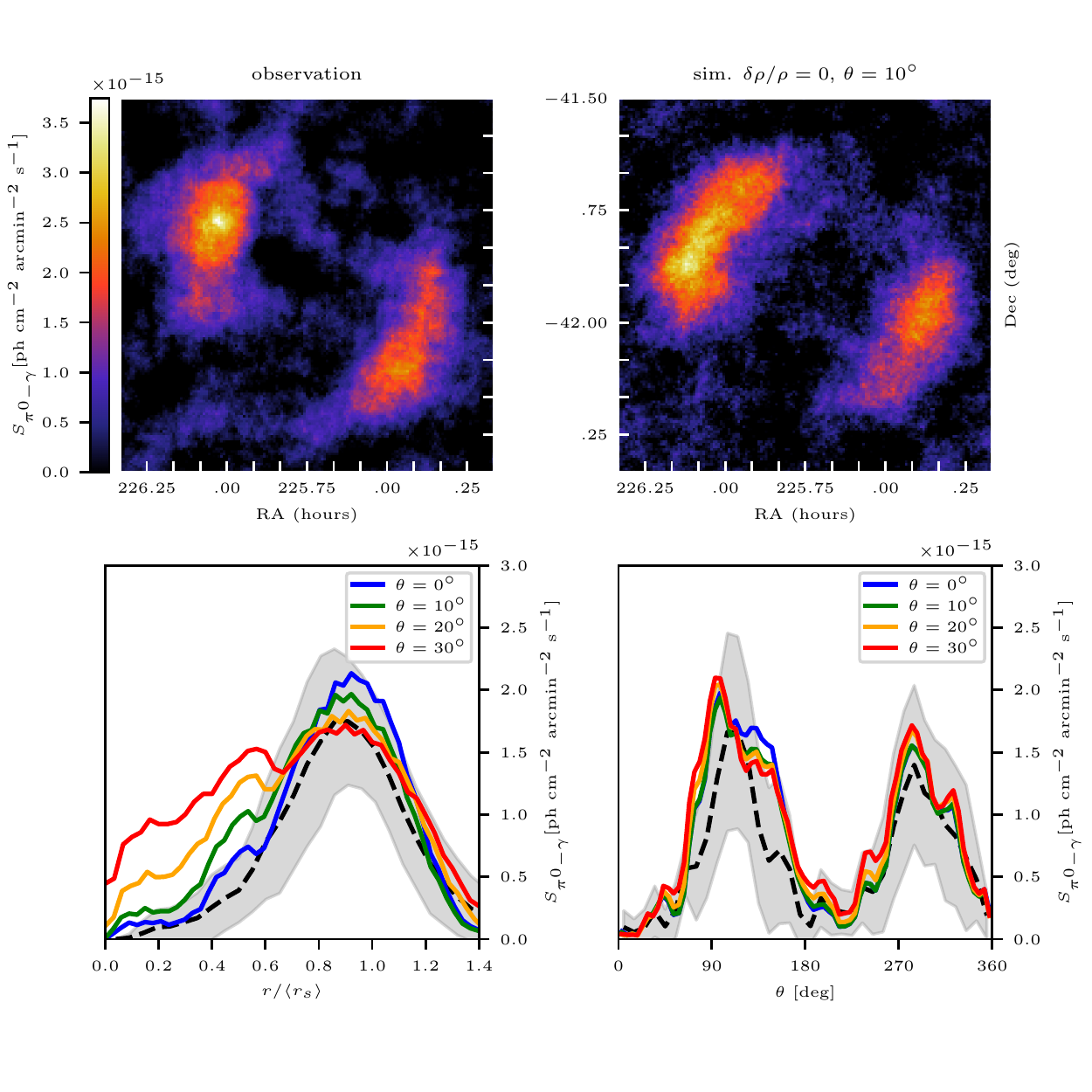}
\caption{Top left: emission map of SN1006 \citep{2010AA...516A..62A}. Top right:
  our best match for SN1006 without density fluctuations and a rotation angle
  $\theta = 10^\circ$. Bottom left: radial profiles of the surface
  brightness. Profiles of the best match for $\theta =
  [0^\circ,10^\circ,20^\circ,30^\circ]$ are reported for comparison with the
  observed one (black dashed line). The grey-filled areas represent the standard
  deviations of the observed radial profiles, respectively. The more strongly
  inclined cases ($\theta=20^\circ$ and $\theta = 30^\circ$) fail to reproduce
  the declining emission inside and outside the maximum. Bottom right:
  Comparison of simulated azimuthal profiles of SN1006 and observed data (black
  dashed line) with standard deviation (grey). While there is broad agreement
  between the models and observation, the NE cap (at around 90$^\circ$) is
  narrower in the HESS observation.}
\label{Fig: SN1006_best_match}
\end{figure*}

\section{Morphological modelling of SN 1006}
\label{sec: SN1006}
In the following two sections, we use our intuition developed in our parameter
study in Fig.~\ref{Fig: 12_plots} to find the most promising combination of
magnetic and density inhomogeneities in the obliquity-dependent acceleration
scenario to mock the observed gamma-ray emission morphology of shell-type SNRs.
To this end, we analyse the bi-lobed morphology of Type Ia SN1006 in this
section and present our analysis of the two shell-type core-collapse SNRs
Vela Jr.\ and RX J1713 in Section~\ref{sec:CC_SNRs}.

The gamma-ray excess map of SN1006 reported by the \citet{2010AA...516A..62A}
strongly correlates with the synchrotron X-ray emission map and suggests an
emitting region compatible with a thin shell. The peculiar polar cap geometry in
the emission is also observed in other wavelengths such synchrotron X-ray
emission, indicating an acceleration process compatible with efficient
quasi-parallel shock acceleration \citep{Winner}.  Polarimetric radio
observations of the limbs strongly suggests that the ambient field is aligned
along the SE–NW direction \citep{2013AJ....145..104R}, as later confirmed by
recent theoretical models \citep{2015MNRAS.449...88S}. Moreover in
\citet{2020MNRAS.496.2448P} is shown that the correlation length of the magnetic
field in case of pure turbulence is at least 15 times the angular size of the
SNR and consistent with a homogeneous field across the SNR.

All these models used a setup involving a homogeneous ISM. However, if the TeV
gamma-ray emission is mainly from hadronic sources, the brighter NE lobe
suggests the presence of a large scale density gradient pointing from SW to
NE. Furthermore, the small-scale brightness variations at the outer shock radius
could either be caused by (i) density inhomogeneities in the ISM that would
corrugate the shock upon colliding with these inhomogeneous structures or (ii)
by obliquity-dependent shock acceleration in combination with small-scale
turbulent magnetic field superposed on a homogeneous magnetic field. We will
study the impact of both effects on the gamma-ray surface brightness maps in the
presence of a large-scale density gradient.

\subsection{Simulation model}

 We proceed with a suite of simulations with the same
large-scale density gradient but with different setups for the density and
magnetic field topology. We conducted a number of exploratory simulations to
test the steepness of the gradient in order to faithfully reproduce the
different TeV gamma-ray brightness of the two lobes and match the measured
fluxes reported in \citet{2010AA...516A..62A}.  For the average value of the
density and SNR distance we used the same values reported by the
\citet{2020MNRAS.496.2448P}. With reference to Eq.~\ref{eq: density} we used
$\psi = 135^\circ$ for the orientation and $\Delta = 0.9$ for the density
slope. In physical units this corresponds to a density gradient of $\nabla n
\simeq 0.0034~\cm^{-3}~\pc^{-1}$ pointing from SW to NE, which means that in the
NE region the average ISM density is $\langle n \rangle \sim 0.12~\cm^{-3}$
while in the SW rim we find $\langle n \rangle \sim 0.08~\cm^{-3}$. Initially, we
adopt a homogeneous magnetic field 5 $\muG$ throughout the simulation domain.

Our baseline model does not adopt turbulent density and magnetic
fluctuations. Our second model assumes turbulent density fluctuations with a
amplitude $\delta\rho/\rho\sim0.5$ and coherence scale
$\lambda_\rho=5~\pc$. Finally, our third model has no turbulent density
fluctuations, but adopts magnetic turbulence with $f_B\sim0.3$ and coherence
scale $\lambda_B=5~\pc$.  We simulated and evolved our SN1006 model for $1010$
yrs, constructed the pion-decay gamma-ray surface brightness map resulting from
hadronic CR interactions and convolved it to the observational PSF of width
$\sigma = R_{68}/1.515 = 0.042^\circ$, with $R_{68}= 0.064^\circ$
\citep{2010AA...516A..62A}. All other parameters used for the simulations are
reported in Table~\ref{table:Table 1}.

\subsection{Noise modelling}
In order to disentangle physical effects that generate small-scale brightness
fluctuations from spurious instrumental effects, we need to accurately model the
instrumental noise. This noise dominates the lower excess counts and is
  normally distributed with zero mean.  Here, we explain our procedure of
generating such a noise map for our synthetic gamma-ray emission maps. To this
end, we calculate the noise power spectrum of the excess map of SN1006 and
exclude the emission of the NE and SW lobes. This is done by masking the
original excess map from the \citet{2010AA...516A..62A} with a sharp cutoff
equal to $c_\mathrm{cut} = |c_{\mathrm{min}}|$ where $c_{\mathrm{min}} = -20$ is
the minimum value of the excess counts of the H.E.S.S. observations of
SN1006. This procedure is justified by the fact that the noise excess counts
  follow a normal distribution with zero mean (see Appendix~\ref{sec:app}).

The power spectrum of SN 1006 is obtained via a 2D Fourier transform of the
masked data set.  The noise power spectrum presents two features: (i) a large
scale Gaussian-like noise as expected from the PSF convolution of the
instrument, and (ii) a small scale power law $\propto k^{-2}$ extending to
higher modes. We fit the power spectrum with the following function in
$k$-space:
\begin{equation}
\label{eq: pk}
P(k) = A \exp \left( -\dfrac{k^2}{2 \sigma^2_{k}} \right) + B k^{-2}
\end{equation}
where, $\sigma_k$ represents the standard deviation in $k$-space and the
variables A and B determine the relative strength of the Gaussian and the
power-law tail.  The black dashed line in Fig.~\ref{Fig: power_spectra} shows
the measured noise power spectrum of the SN1006 excess map in comparison to the
simulated noise (blue line). We assume that the noise is fully characterised by
two-point correlations and obtain a random realisation of the noise via 2D
inverse Fourier transform.  This noise map is then superposed on the mock
gamma-ray emission map that was convolved with the observational point spread
function (PSF).

\subsection{Simulated TeV emission}

We show the resulting synthetic maps of SN 1006 in comparison to the original
excess map in Fig.~\ref{Fig: SN1006_comparison}. We report three gradient models
with the following assumptions (from left to right): (i) no density
fluctuations, (ii) density fluctuations with $\delta\rho/\rho=0.5$ and no
magnetic field fluctuations, and (iii) magnetic field fluctuations with
$\delta B/B=0.3$ and no density inhomogeneities. This enables us to separately
analyse the effect of these configurations on the morphology of SN1006.

While our noise modelling is responsible for emission in regions where the
acceleration is supposed to be extremely inefficient or absent, the density
perturbations are able to corrugate and smooth the boundaries of the lobes, in
particular of the SW lobe for the chosen seed. The resulting picture appears
with a defined bi-lobed structure and an increasing level of turbulence in the
medium causes the emergence of secondary morphological details.  We notice that
density perturbation of $\delta\rho/\rho=0.5$ (third picture of Fig.~\ref{Fig:
  SN1006_comparison}) affect the morphology of the SW lobe resulting in a
fainter surface brightness and a broader shape.  This suggests that our
preferred model for SN 1006 in the TeV gamma-ray band is given by a model with
homogeneous density.

Similarly, we observe the emergence of secondary morphological details from a
moderate level of turbulence in the magnetic field (fourth picture of
Fig.~\ref{Fig: SN1006_comparison}) in our obliquity dependent CR acceleration
scenario. We compare this model with a coherence scale at $\sim 1/3$ the size of
the remnant to our other two models with a homogeneous density and a moderate
level of density inhomogeneities. In spite of the low level of magnetic
fluctuations ($\delta B/B=0.3$) the morphology significantly deviates from the
observed bi-lobed shell-type morphology of SN1006 and shows significant
substructures. In addition, this model is also ruled out by the excess of the
emission in the central region of the remnant which contrasts to the clean
separation of the two lobes as observed by the H.E.S.S.\ collaboration. This
sets an approximate upper limit for the local turbulence of the magnetic field
in a obliquity dependent scenario.

Another degree of freedom is the orientation of the homogeneous magnetic field
$\vect{B}_0$ with respect to the line of sight. To study its influence on the
emission map and (radial and azimuthal) profiles, we perform a three-dimensional
(3D) rotation of the SNR around the axis perpendicular to the orientation of
$\vect{B}_0$ into the line of sight.  We show different inclinations ($\theta =
[0^\circ,10^\circ,20^\circ,30^\circ]$) of $\vect{B}_0$ with respect to the plane
of the sky, similarly to \citet{2011A&A...531A.129B} for the X-ray emission to
constrain the 3D orientation of the magnetic field of SN 1006. Note that this is
somewhat degenerate with the assumed 3D orientation of the density gradient that
we also assume to lie in the plane of the sky.

An increasing inclination results in a less pronounced peak and a broader
distribution of gamma rays as the hadronic emission from the efficiently
accelerated CRs at the quasi-parallel shocks changes from being limb-brightened
to contributing to a broader solid angle on the sky.  The top right panel in
Fig.~\ref{Fig: SN1006_best_match} shows our best match for SN1006 after the
aforementioned rotation of the SNR. Indeed, the lobes appear slightly more
diffuse, predominantly in the internal regions.

To quantify the effect of magnetic inclination we compute the radial and azimuthal
profiles of our synthetic maps according to the sketch shown on the right-hand
side of Fig.~\ref{Fig: power_spectra}. The radial profiles are computed within
the quadrants where quasi-parallel shock acceleration takes place while the
azimuthal profiles show an average in the radial range $2/3 < r/\langle
r_\rmn{s} \rangle < 4/3$, where $r_\rmn{s}$ is the average shock radius of the
SNR.

The comparison of the post-rotation radial profiles to the observed radial
profile of SN 1006 provides an important constraint for the maximum magnetic
inclination. These profiles are show in the bottom panels of Fig.~\ref{Fig:
  SN1006_best_match}.  We include a $1 \sigma$ standard deviation grey-filled area to the
observed SN1006 profiles. While larger inclinations
($\theta>10^\circ$, orange and red lines) result in emission profiles that are
too extended, an inclination of $\theta\leq10^\circ$ is in better agreement with
the observed radial profile. Barring our remark about the orientation of the
density gradient, we conclude that the magnetic inclination does not
exceed $10^\circ$ with respect to the plane of the sky.

A minor element of discordance can be also found in the azimuthal profiles
(Fig.~\ref{Fig: SN1006_best_match}, bottom right) for different rotation
angles. While the simulated SW lobes coincide with the observed one in relative
brightness and extension, the simulated NE lobe is slightly broader in
comparison to the observed lobe.  This suggests that the critical angle for the
magnetic obliquity may be smaller than $45^\circ$ inferred by recent PIC
simulations.  Regardless the spatial rotation all the models reproduce the
observed modulation of the two lobes within the 1-$\sigma$ uncertainty.

\section{Morphological modelling of core collapse SNRs}
\label{sec:CC_SNRs}
We now turn to two other well-known SNRs with a clear shell-type morphology:
Vela Jr. and RX J1713. Both SNRs are of core-collapse origin which means that
the ambient ISM is a star-forming region of multi-phase, turbulent gas. We
account for this by simulating the supernova explosion in a multiphase ISM with
a large population of small, dense gaseous clumps with a typical overdensity of
$10^4$ in comparison to the ambient ISM and adopt a purely turbulent magnetic
field without a mean field, $\vect{B}_0=\mathbf{0}$.

\subsection{Observational constraints on Vela Jr.}
\label{sec: VelaJr}

The \citet{2018AA...612A...7H} has observed TeV gamma-ray emission from the SNR
Vela Jr. (RX-J0852.0-4622) with a resolved gamma-ray spectrum. Estimates on the
SNR age vary from a very young remnant of $\sim 700$ yrs
\citep{1999A&A...350..997A} to an older object of more than 5000~yrs
\citep{2008ApJ...678L..35K}.  The SNR can be a nearby object at $D = 0.2~\kpc$,
as inferred from studies of the decay of ${}^{44}\mathrm{Ti}$ nuclei
\citep{1998Natur.396..142I}, or a more distant one at $D = 0.75~\kpc$, as
inferred from the slow expansion of X-ray filaments \citep{2008ApJ...678L..35K}.
The presence of interstellar molecular clouds suggests that the origin of
TeV-gamma rays from these objects is mainly hadronic
\citep{2013ASSP...34..249F}.

The lack of thermal X-ray emission places a very low limit at $n=0.03~\cm^{-3}$
while assuming a homogeneous environmental density
\citep{2001ApJ...548..814S}. However, if the ISM is composed of dense clumps
that are embedded in a lower-density hot ambient phase, the resulting thermal
X-ray emission (of the hot phase) is lower while the presence of the dense
clumps implies a higher average density. A conventional approach in the hadronic
model is to use a density of the order of $n \sim 1~\cm^{-3}$
\citep{2006A&A...449..223A}, while hydrodynamic models suggest values of less
than $0.4~\cm^{-3}$ \citep{2015ApJ...798...82A}.  More recently HI and CO
measurements and partial morphological correspondence with the TeV-gamma ray
morphology indirectly suggest an extremely high average ISM density of the order
of $n \sim 100~\cm^{-3}$ \citep{2017ApJ...850...71F}.  However there is no
direct observational evidence that the clumped gas is in direct physical contact
with the shock-accelerated cosmic rays. In fact the passage of the shock
dissipates kinetic energy, heats the ions to particle energies of several keV
and is directly responsible of dissociation of CO and $\mathrm{H}_2$ and the
ionisation of the neutral part of the clouds on a time-scale of the order of a
few years \citep{2019MNRAS.487.3199C}.

As the shock overruns the magnetised ISM, magnetic fields are draped around the
dense clouds, precluding the diffusion of cosmic rays deep into the cloud so
that the TeV cosmic rays can only probe a narrow skin of the cloud. The
penetration depth of this skin can be estimated by realising that the draped
magnetic field reaches strength of order $B\approx\sqrt{8 \pi \alpha \rho v^2}
\approx1\mathrm{mG}$ \citep{2008ApJ...677..993D, 2010NatPh...6..520P} for
typical parameters $\alpha=2$, $n= 0.1~\cm^{-3}$, and
$v_\mathrm{s}=3000~\km~\s^{-1}$. If the average ISM density were indeed
$100~\cm^{-3}$ \citep{2017ApJ...850...71F}, this should yield draped magnetic
field strengths of 32~mG, which should be observable via Zeeman splitting of
which there is no evidence which argues against such a high average density. The
skin depth of the dense cloud reachable by TeV CRs, assuming Bohm diffusion,
varies from a few to several gyro radii:
\begin{equation}
r_g = \dfrac{p_\perp c}{e B} \approx 10^{-6}~\pc \left( \dfrac{pc}{\TeV} \right) \left(\dfrac{B}{\mG}\right)^{-1}
\end{equation}
which is negligible in comparison of the cloud size (0.1~pc). Even if the
magnetic wrap is not perfect and CRs can penetrate 100 gyro radii inside the
cloud, then the fraction of the cloud volume seen by the TeV CRs is $\Delta
V\sim1 - (1-10^{-3})^3 \sim 3\times10^{-3}$. Our assumed total dense cloud
mass of $45~\rmn{M}_\odot$ that is physically associated with the SNR is a
fraction of $2\times10^{-3}$ of the available gas mass of
$2.5\times10^4~\rmn{M}_\odot$ \citep{2017ApJ...850...71F} towards the Vela
Jr.\ region, some of which may be projected onto the SNR but is not physically
associated to it and only a tiny fraction of the molecular and neutral gas
inside the Vela Jr. SNR is seen by cosmic rays due to magnetic draping,
suppressing cosmic ray propagation into the cloud before the cold gas gets
ionised and dissociated. These considerations justify our assumption of only
accounting for CR advection in our simulations.

In addition, not all the mass of the clumps engulfed by the shock has been
processed by the shock because of the strong deceleration of the blast wave
inside the clumps (see Appendix of \citealt{2020MNRAS.496.2448P}).  Combining
these arguments suggests that only a tiny fraction of the dense
(neutral/molecular) phase of the ISM is in physical contact with the shock.  For
the magnetic field we decided to follow the same prescription used in
\citet{2020MNRAS.496.2448P} setting the coherence scale of the turbulent
magnetic field to $\lambda_B = 13~\pc$ and adopt $f_B=1$.  The entire set of
parameters used in the simulation is summarised in Table~\ref{table:Table 1}.

\subsection{Observational constraints on RX J1713}
\label{sec: RX J1713}

RX J1713 represents another bright TeV-emitter with a distinct shell-like
emission morphology. This SNR is subject to intense studies thanks to its strong
non-thermal X-ray emission and the detection of high-energy and very-high-energy
gamma-rays. \citet{1997A&A...318L..59W} suggested that RX J1713 is linked to an
AD393 guest star which, according to historical records, appeared in the tail of
constellation Scorpius, close to the actual position of the remnant. This would
put the age of the remnant close to $1.6~\kyr$. More recent estimates based on
X-ray emission combined with hydro models in homogeneous media suggest an older
remnant age of $6.8^{+1.1}_{-2}\kyr$ \citep{2020ApJS..248...16L}.

The distance of the object is estimated to be around $1~\kpc$ \citep{2003PASJ...55L..61F} while  
\citet{2008ApJ...685..988T} suggests a distance interval between 0.5~kpc and 1~kpc. 
Assuming a free expansion model this translates into an upper limit of the average 
shock speed of about $\langle v_\mathrm{s}\rangle = 6300~\km~\s^{-1}$. Proper motions 
of bright X-ray filaments instead place the shock velocity between $\sim 1000~\km~\s^{-1}$ 
and $\sim 4000~\km~\s^{-1}$ \citep{2003PASJ...55L..61F}.

A hadronic origin of the TeV gamma-ray emission in RX J1713 was suggested in
several papers \citep{2010ApJ...708..965Z,2014MNRAS.445L..70G}. The distribution
of gas in RX J1713 is crucial to establish the origin of the observed
gamma-rays.  An upper limit for the diffuse density of $<2~\cm^{-3}$ is derived
from non thermal X-rays measurements \citep{2008PASJ...60S.131T}. To explain the
lack of thermal X-ray emission \citet{2004A&A...427..199C} sets an even lower
upper limit for the ambient density of $n\sim 0.02~\cm^{-3}$. However, this
argumentation can be circumvented by introducing a densely clumped environment.

Studies of non-thermal X-ray emission, TeV gamma-ray emission and their partial
spatial correlation with HI ans H2 suggest the presence of numerous dense clumps
in the ISM \citep{2003PASJ...55L..61F, 2015ApJ...799..175S,
  2009atnf.prop.2223R}.  However the same argument used for Vela Jr. applies:
the lack of direct evidence of physical contact between the gas distribution and
the shock and the subsequent ionisation of HI casts doubts on its physical
association with the SNR.  The interaction of a blast wave with interstellar
clouds and its application to RX J1713 has been studied in
\citet{2012ApJ...744...71I} and applied in \citet{2019MNRAS.487.3199C} for a
shock wave with constant velocity interacting with a target mass of
$45~M_\odot$.

Recently, \citet{2016PASJ...68..108T} calculated the evolution of RX J1713 in
various scenarios such as the case of an expansion in a wind-blown cavity with a
typical density profile $\rho \propto r^{-2}$ and a free expansion dominated by
the ejecta ($\rho_\mathrm{ej} \propto r^{-n}$ with $n=7$). Here we consider
simple Sedov-Taylor expansion models which represent the upper limit for the
expansion of a SNR in homogeneous media \citep{1999ApJS..120..299T}.

To reproduce the saturated NW rim of the remnant we applied a homogeneous
positive large-scale gradient to the initial conditions pointing from SE to NW
with $\Delta=1.6$, corresponding to a density gradient of $\nabla n =
0.02~\cm^{-3}~\pc^{-1}$.  Superposed on this density gradient we insert a
population of dense clumps of size $0.1~\pc$ using exactly the same setup as for
Vela Jr.  For the magnetic field we selected a fully turbulent setup ($f_B=1$)
with a coherence length of $\lambda_B = 13~\pc$, not too dissimilar from the
size of the remnant ($\sim 17.4~\pc$).  The entire set of parameters used in the
simulation is summarised in Table~\ref{table:Table 1}.

\begin{figure*}
\centering\includegraphics[width=\textwidth]{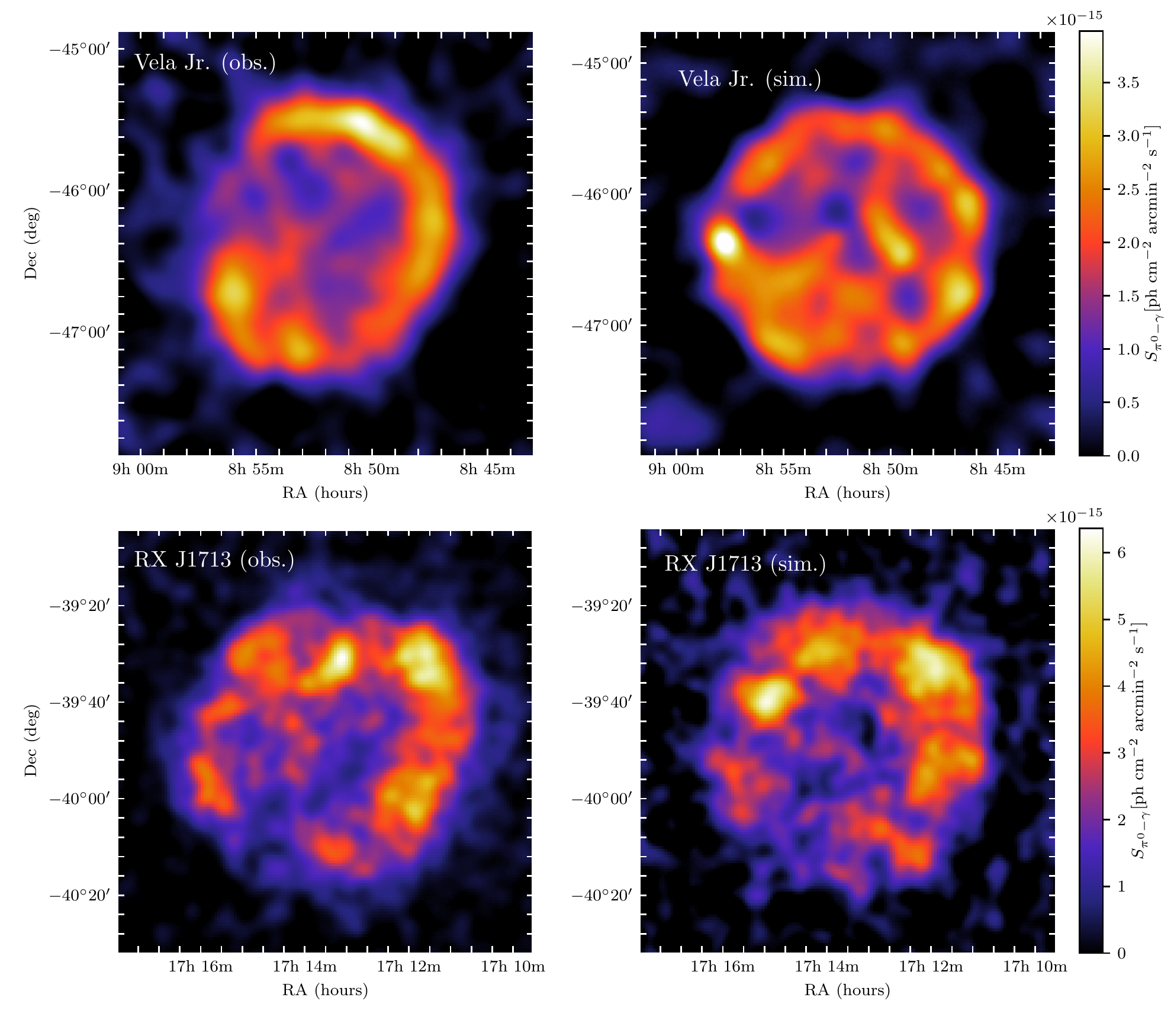} 
\caption{Comparison between observed gamma-ray emission maps (left) of Vela
  Jr. \citep{2018AA...612A...7H} and RX J1713 \citep{2018A&A...612A...6H} and simulated gamma-ray maps (right)
  of the SNRs for our model with obliquity-dependent CR acceleration, a
  turbulent magnetic field with $f_B=1$ and $\lambda_B=13$ and additionally a
    gradient for RX J1713 with $\Delta=1.6$. In our simulated maps we add
  Gaussian noise with the observed power spectrum and use the PSFs appropriate
  to each observation ($0.08^\circ$ and $0.036^\circ$ for Vela Jr. and RX J1713,
  respectively). We also rotated mock emission maps of the SNRs to match the
  azimuthal position of the faintest region in the shell.}
\label{Fig: 2_SNRs}
\end{figure*}
\begin{figure*}
\centering\includegraphics[width=\textwidth]{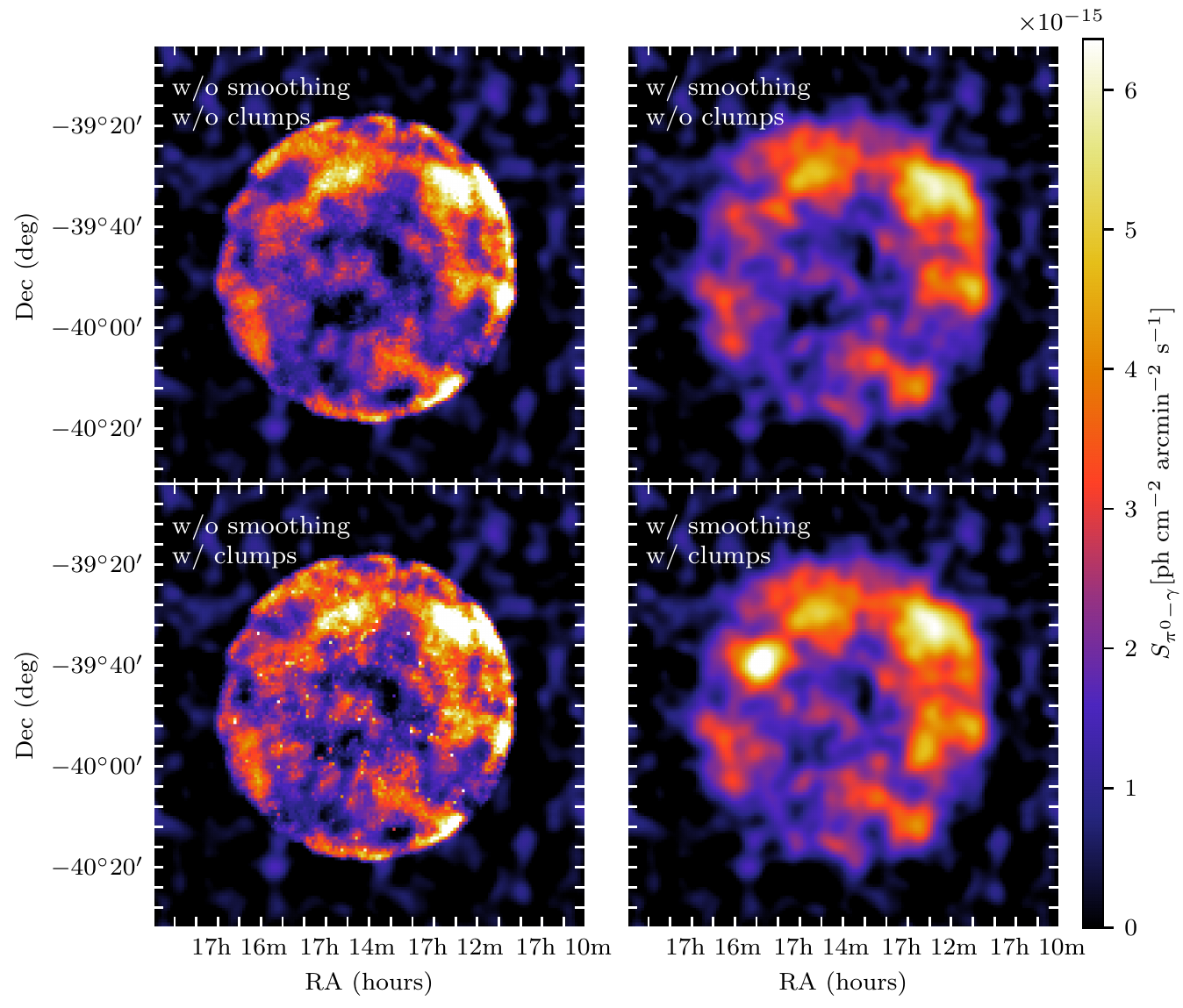} 
\caption{Comparison between the simulated surface brightness maps of RX J1713
  without clumps (top row) and with clumps (bottom row). The comparison between
  the maps without (left) and with PSF smoothing (right) shows the contribution
  of the clumps in specific portions of the outer shell and in the central
  region. While most of the gamma-ray substructure is due to obliquity-dependent
  CR acceleration, individual bright patches in the map can be due to the
  hadronic emission of dense clumps.}
\label{Fig: RX_4_Plots}
\end{figure*}
\begin{figure*}
\centering\includegraphics[width=\textwidth]{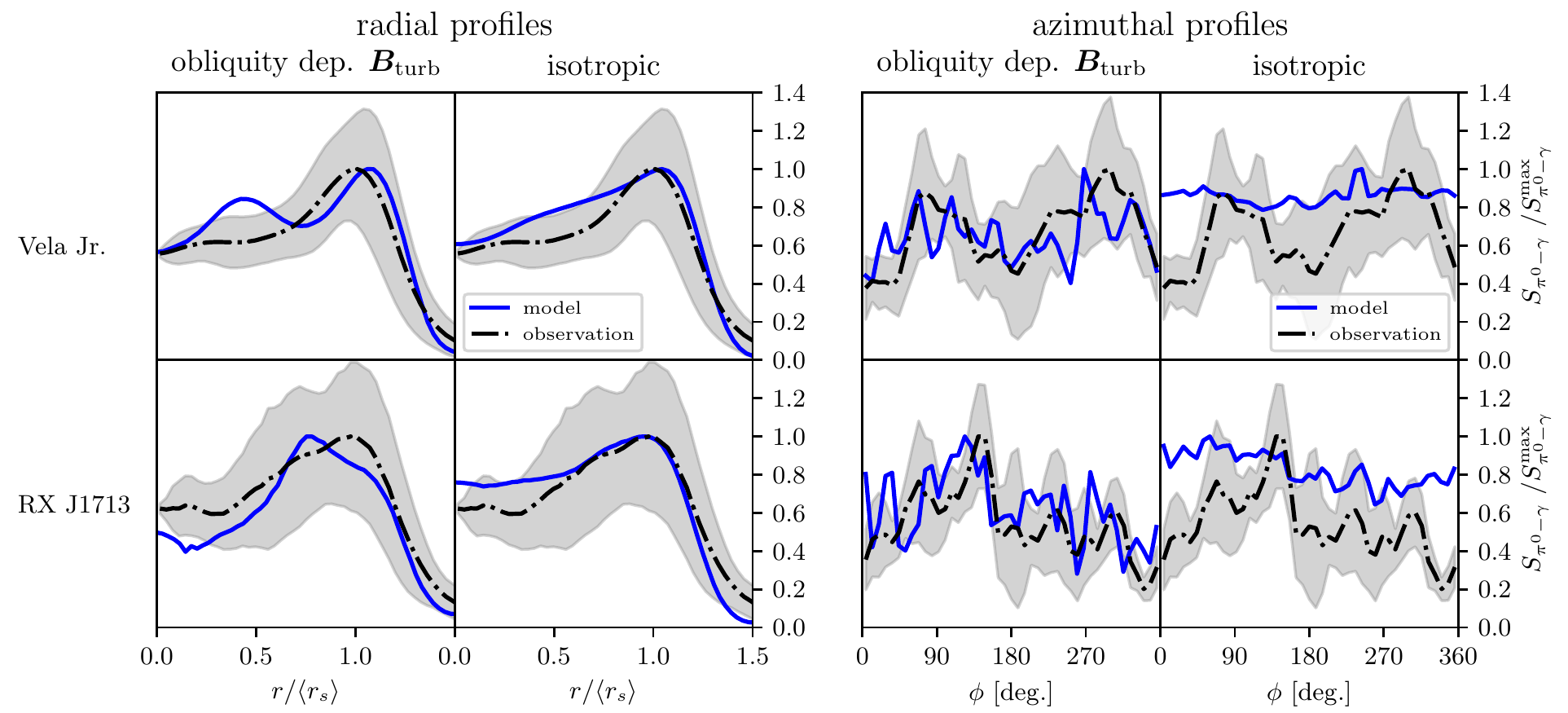} 
\caption{Normalised radial (left-hand side) and angular (right-hand side)
  profiles of the models (blue lines) shown in Fig.~\ref{Fig: 2_SNRs} compared
  to SNRs Vela Jr.\ (dashed dotted lines in the top row,
  \citealt{2018AA...612A...7H}) and RX J1713 (dashed dotted lines in the bottom
  row, \citealt{2018A&A...612A...6H}). The grey-filled area represents the
  $1\sigma$ uncertainty for both SNRs. The column labelled with ``obliquity
  dep.\ $\vect{B}_{\mathrm{turb}}$'' shows our obliquity-dependent acceleration
  models for a turbulent magnetic field while the label ``isotropic'' refers to
  our isotropic acceleration model.  While both models reproduce the radial
  emission profiles well, the isotropic model clearly fails to capture the
  azimuthal brightness variations seen in the data which is statistically
  consistent with our obliquity-dependent acceleration model. }
\label{Fig: 2_SNRs_radial_angular}
\end{figure*}

\subsection{Simulations}

Figure~\ref{Fig: 2_SNRs} shows the comparison between the observed excess
brightness maps of Vela Jr. and RX J1713 (left-hand side) and our PSF-convolved
synthetic maps derived from turbulent models with density inhomogeneities and
dense clumps (right-hand side). To model the noise in the post-processing we
applied the same method used for SN1006.  In the final step we rotate the
surface brightness maps in order to approximately match the low and high
emissivity outer shell regions of the observed maps. Our simulation models
provide a good agreement with the data.

The presence of high-density molecular clumps provides an important contribution
to the global emission morphology. The effect of the clumps on RX J1713 is shown
in Fig.~\ref{Fig: RX_4_Plots}. While we notice that the orientation of the
magnetic field is mainly responsible of the patchy morphology of the outer shell
the clumps add several bright spots to the map without modifying the expansion
history of the SNR due to their negligible volume filling factor. A comparison
with the left-hand side figure in the bottom row shows that the strong smoothing
applied to the synthetic maps makes it impossible to resolve the emission
originating from a single isolated clump and that clusters of bright clumps are
similar to large magnetic field patches oriented quasi-parallel to the shock
normal, thus enabling efficient CR acceleration. 

The effect of obliquity-dependent shock acceleration is better shown in
Fig.~\ref{Fig: 2_SNRs_radial_angular}, where we compare radial and azimuthal
profiles of our simulated gamma-ray maps to those of the excess maps of Vela
Jr.\ and RX J1713. We consider the two cases of (i) obliquity-dependent shock
acceleration in a turbulent magnetic field and (ii) isotropic CR acceleration, both
for a constant ambient density and and global density gradient. While the pure
isotropic models (without clumps and PSF smoothing) show and enhance level of
surface brightness in the central region compared to the obliquity dependent
acceleration models (see Fig.~\ref{Fig: 12_plots}), the inclusion of dense
clumps and convolution with the observational PSF fills in the central parts to
similar emission levels.

Most notably, the isotropic acceleration models cannot reproduce the observed
azimuthal small-scale variations in the surface brightness as shown in the
right-hand panels of Fig.~\ref{Fig: 2_SNRs_radial_angular}.  Interestingly,
obliquity-dependent acceleration models are able to modulate the emissivity
peaks on a relatively small scale, in a statistically similar fashion. This is a
clear prediction of an obliquity-dependent CR acceleration in a turbulently
magnetised ISM, in which the morphology of the magnetic field, in tandem with
emission from dense clumps, is responsible for the VHE emission morphology
observed by imaging air Cerenkov telescope such as H.E.S.S. and enables us to
infer the magnetic coherence scale of the ISM surrounding the SNR
\citep{2020MNRAS.496.2448P}.  We will address the interesting question whether
large-amplitude density perturbations or extreme density gradients alone are
able to modulate the surface brightness in a similar way to mimic the VHE
observations in Section~\ref{sec: high turbulence}.

\section{The case of a highly turbulent medium}
\label{sec: high turbulence}
Here we study the effect of strong density fluctuations on a supernova explosion
in its early Sedov stage at $t_\age = 1~\kyr$.  We aim at answering two
questions: (i) Can high-amplitude turbulence in the ISM on scales comparable to
or smaller than the size of the remnant mimic a bi-lobed or patchy VHE gamma-ray
emission with strong, small-scale brightness variations observed in the three
SNRs studied here and (ii) can localised strong variations of the density be
responsible for an extreme corrugation of the shock front and its eventual
disruption into smaller clumps, speeding up the end of the Sedov phase, thus
anticipating the beginning of the snowplough phase and the eventual merging of
the fragments with the ISM?  This dynamical effect may feedback on the
acceleration mechanism of CRs.

\begin{figure*}
\includegraphics[width=\textwidth]{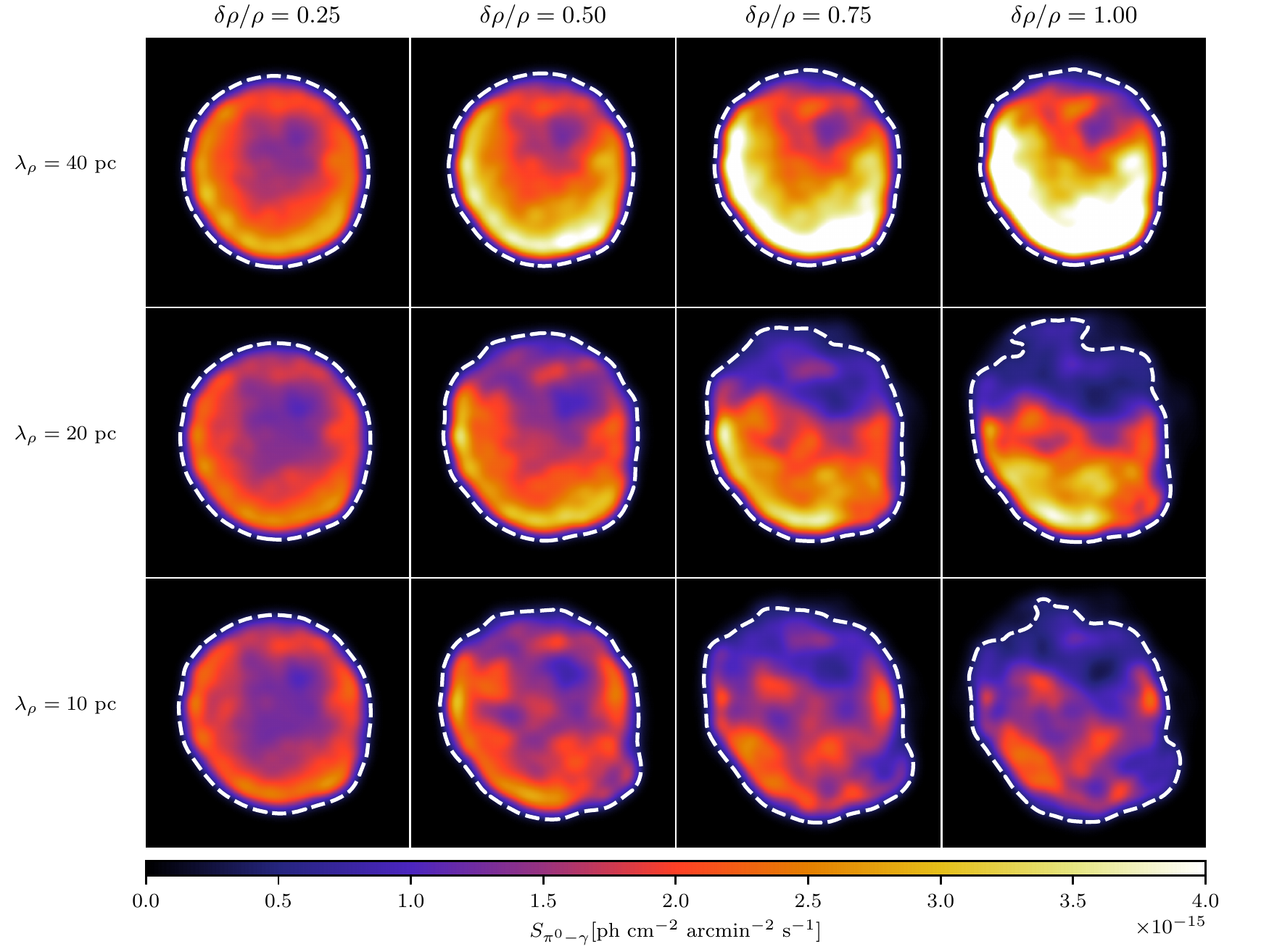} 
\caption{Gamma-ray emissivity from SNRs in the Sedov-Taylor phase with
  large-amplitude density fluctuations in our isotropic acceleration scenario
  that does not depend on magnetic obliquity. From left to right the fluctuation
  strength increases from 0.25 to 1 with respect to the average density in steps
  of 0.25. From top to bottom the correlation length for the density
  fluctuations are, in decreasing order: 40 pc (top row), 20 pc (middle row) and
  10 pc (bottom row). The maps are taken at $t_{\mathrm{age}} = 10^3$ yrs after
  the explosion and all exhibit the same average number density of
  $0.1~\cm^{-3}$. The maps have a side length of $L= 20~\pc$ and are smoothed
  with the PSF of RX J1713 with $\sigma = 0.036^\circ$.  The dashed white
    curves denote the emission contours used in the histograms of Fig.~\ref{Fig:
      radii_distribution} to quantify the degree of corrugation of the outer
    shell boundary of the gamma-ray emission.}
\label{Fig:16_maps}
\end{figure*}
\begin{figure*}
\includegraphics[width=\textwidth]{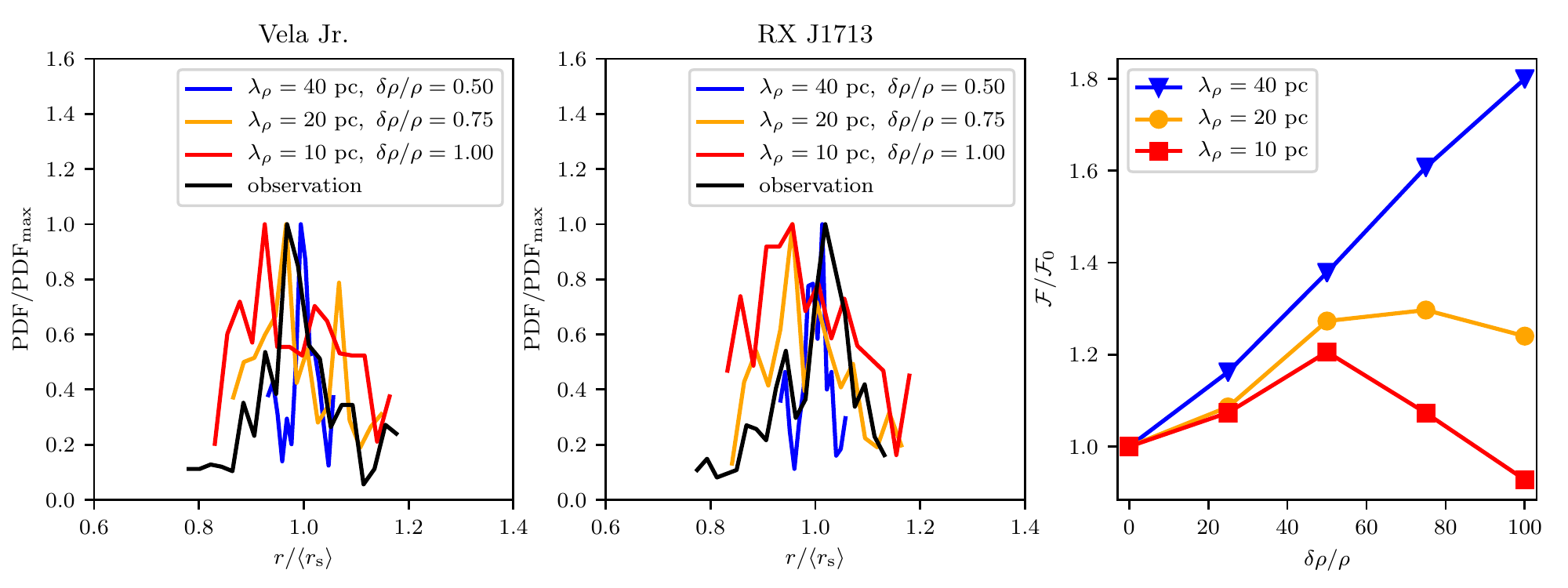} 
\caption{Left and centre: distribution of the radii of the external signal
    contours of the gamma-ray brightness map shown in white in
    Fig.~\ref{Fig:16_maps} for three selected models in our isotropic
  acceleration scenario that does not depend on magnetic obliquity. Increasing
  fluctuation amplitude and correlation length implies a larger dispersion of
  the radial distribution of the outer gamma-ray emission contour and a
  significant departure from spherical symmetry. For comparison we show the
  radial distribution of the SNRs Vela Jr.(left) and RX J1713 (centre). Right:
  Gamma-ray flux of the models pictured in Fig.~\ref{Fig:16_maps} as a function
  of the amplitude of density fluctuations. The flux is normalised to the flux
  $\mathcal{F}_0$ of the model without density turbulence. The plot shows an
  excess emission associated with the coherence length $\lambda_\rho = 40~$pc
  which grows with increasing degree of turbulence.}
\label{Fig: radii_distribution}
\end{figure*}

In order to test whether our core-collapse SNRs necessarily require an
obliquity-dependent acceleration model, we adopt our isotropic CR acceleration
model and systematically vary density fluctuations. To this end, we present a
set of 12 simulations with varying amplitude of turbulence from $\delta \rho /
\rho = 0.25$ to $\delta \rho / \rho = 1$ at steps of $0.25$, and varying
coherence scale of the fluctuations at steps of $L/n$ with box size $L=40~\pc$
and $n=[1,2,4]$.  We use the same setup as in our previous models, which is
described in Section~\ref{sec: methodology}, and show results for the exact same
random seed in Fig.~\ref{Fig:16_maps}. From left to right we present an
increasing turbulent amplitude while the coherence scale of turbulence decreases
from $40$ to $10~\pc$ from top to bottom.

We can clearly see that the shock becomes more corrugated for decreasing
coherence scale while increasing the fluctuation amplitude at constant coherence
scale has a comparably smaller impact on the azimuthal dependence of the shock
propagation speed.  We notice that a coherence length of about the size of the
remnant or smaller has a strong impact on the corrugation of the outer shell. In
particular, the cases of $\lambda_\rho = 10~\pc$ and $\delta \rho / \rho \geq
0.75$ (two bottom right panels of Fig.~\ref{Fig:16_maps}) show a disruption of
the shock front at two locations corresponding to extremely under-dense
regions. The breaking of the shell corresponds also to a lower global gamma-ray
surface brightness, signalling that less material is accelerated and that
eventually the shock escapes detection in that region.

Hence the appearance of a bi-lobed structure as in SN1006 would require extreme
fine-tuning of the density distribution which clearly rules out this isotropic
CR acceleration scenario in this case. Our parameter study presented in
Fig.~\ref{Fig:16_maps} shows that in order to obtain significant surface
brightness variations as is observed in the core-collapse SNRs RX J1713 and Vela
Jr., the level of density fluctuations needs to be significant, with $\delta
\rho / \rho \gtrsim 0.75$. However, this implies a heavily corrugated shock
surface which appears to be in conflict with the overall spherical appearance of
SNRs RX J1713 and Vela Jr.

This observation is separately quantified for both SNRs by computing the
  histogram of the radii of the external signal contours of the TeV gamma-ray
  emission maps (see Fig.~\ref{Fig:16_maps}). First, we identify the centroid of
  the SNRs and---in the case of the observations---exclude the background
  noise. We then determine the external emission contour for both remnants by
  choosing a threshold of 10 per cent of the maximum excess counts. We show the
  histogram of these radii, normalised by an average radius $\langle
  r_\mathrm{s}\rangle$ for both remnants, respectively, in Fig.~\ref{Fig:
    radii_distribution}.  We compare those observed radial emission profiles to
the profiles of our simulation model (assuming isotropic CR acceleration) for
$\lambda_\rho = 40~\pc$ with $\delta \rho /\rho = 0.5$ (blue line), for
$\lambda_\rho = 20~\pc$ with $\delta \rho / \rho = 0.75$ (orange line), and for
$\lambda_\rho = 10~\pc$ with $\delta \rho / \rho = 1$ (red line).  While large
scale fluctuations with $\lambda_\rho = 40~\pc$ show a moderate dispersion
despite the high level of density fluctuations, the models with $\lambda_\rho
\leq 20~\pc$ are not compatible with the well defined radial dispersion of Vela
Jr. and RX J1713. This allows to constrain the level of large-scale density
fluctuations to be less than $75$ per cent and $\lambda_\rho > 20~\pc$. The
resulting gamma-ray patchiness is thus not any more strong enough to explain the
small-scale gamma-ray brightness variations observed in the SNRs RX J1713 and
Vela Jr., which thus rules out this isotropic CR acceleration scenario in these
SNRs as well and favours the obliquity-dependent CR acceleration scenario in all
shell-type SNRs studied.

\begin{figure}
\includegraphics[width=\columnwidth]{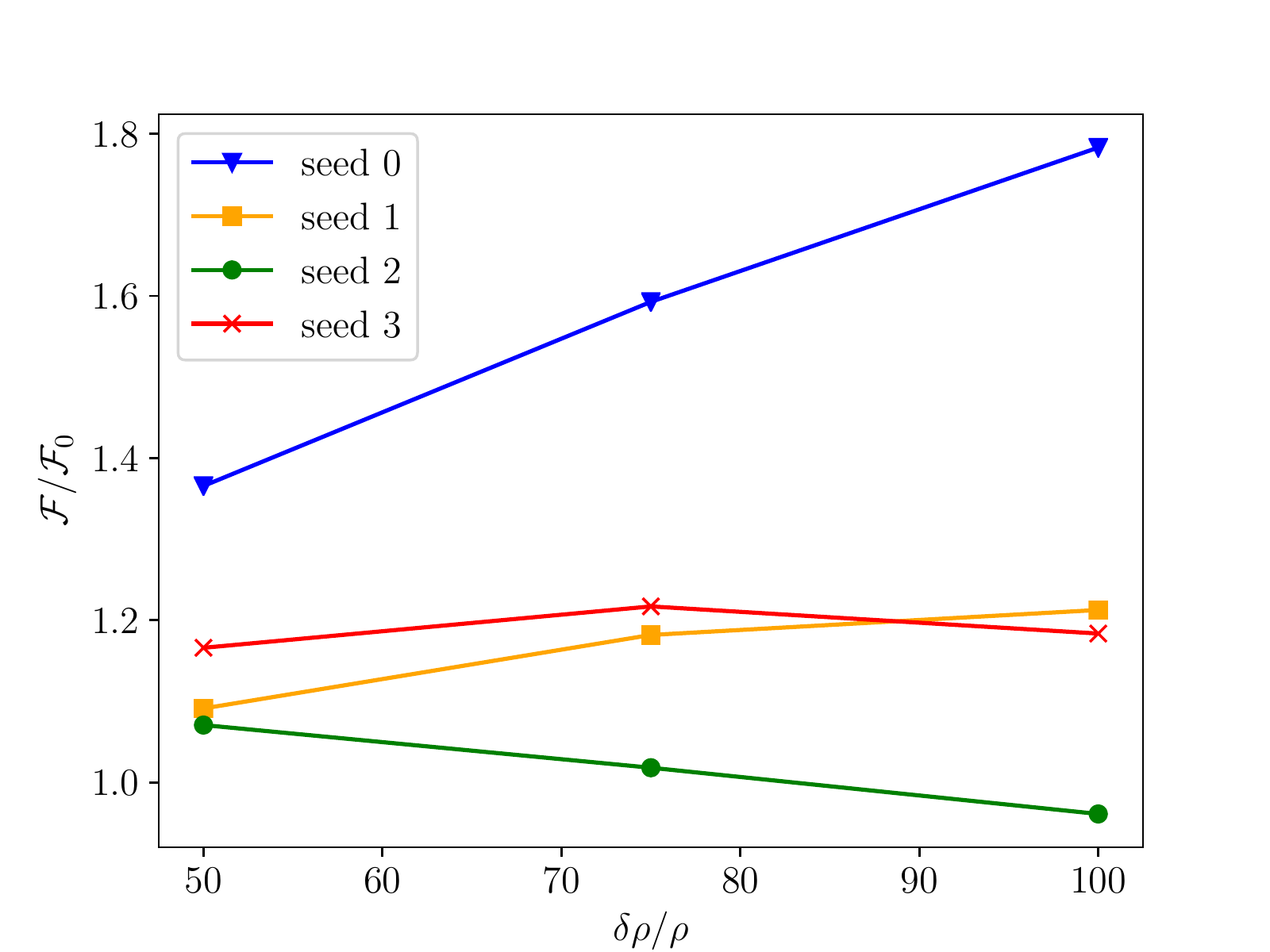} 
\caption{Gamma-ray flux for our turbulent density distribution with
  $\lambda_\rho=40~\pc$ for different random realisations of turbulence as
  function of the turbulent amplitude. The seed labelled with 0 represents the
  one used for the realisations in Fig.~\ref{Fig:16_maps}.  The plot shows no
  clear trend associated with the turbulent amplitude. Random seed 0 causes a
  larger gamma-ray intensity because of the specifics of the overdensities which
  lead to the formation of a bright shell while this effect is not significant
  for the other tested random seeds.}
\label{Fig: flux2}
\end{figure}

The maps with a high level of turbulence and a small coherence scale have a
fainter surface brightness. To quantify the lower gamma-ray efficiency of
small-scale highly turbulent SNRs, we show the fluxes of our 16 models as a
function of the degree of turbulence and of the coherence scale in the
right-hand panel of Fig.~\ref{Fig: radii_distribution}. We find a striking
increase in flux with increase turbulent amplitude for our model with
$\lambda_\rho = 40~\pc$. On the contrary for $\lambda_\rho \leq 20~$pc the
opposite is true. In order to check whether this is due to a particular random
realisation of our turbulent density field or a systematic effect, we simulate
the case of $\lambda_\rho=40~\pc$ with $\delta \rho/ \rho = [0.5, 0.75, 1]$
with three different random realisation of turbulence and show the results in
Fig.~\ref{Fig: flux2}.  The different monotonic and non-monotonic behaviour of
the gamma-ray flux for each random realisation indicates that this behaviour is
not systematic and due to random variance.

\section{Spectra}
\label{sec: spectra}

\begin{figure*}
\includegraphics[width=\textwidth]{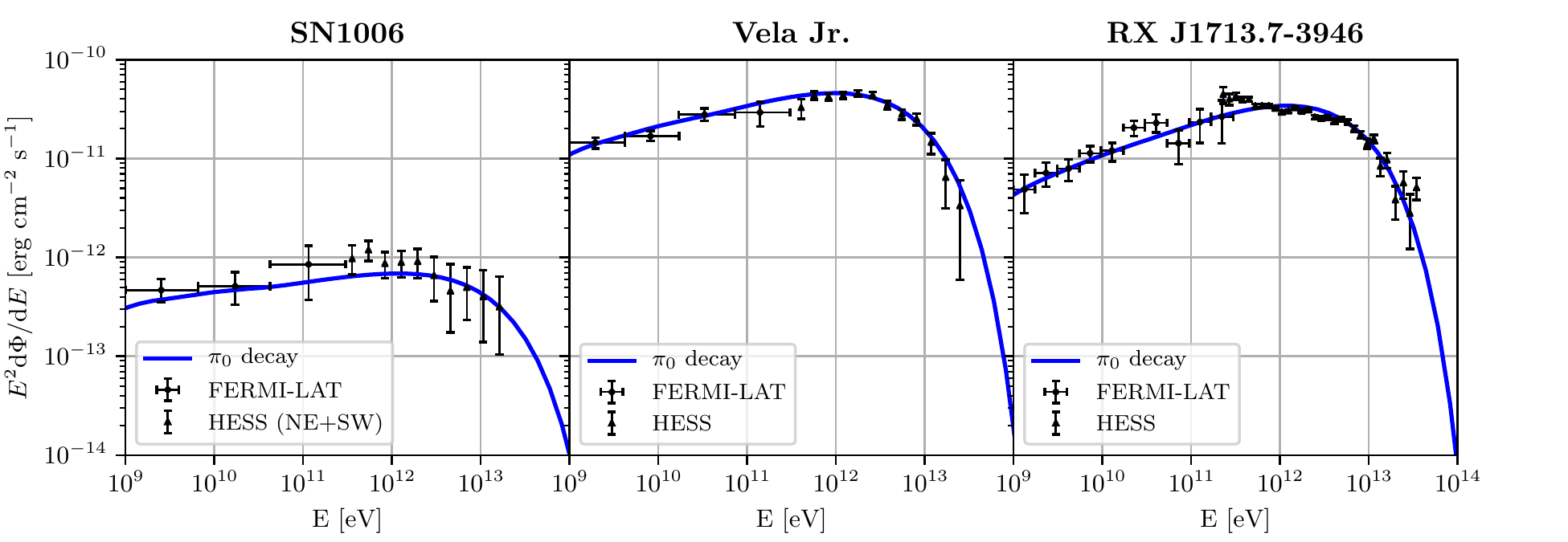} 
\caption{High energy spectra spectra of SN1006 (left), Vela Jr.\ (centre) and RX
  J1713 (right). The models assume a hadronic pion-decay emission scenario for
  the three SNRs. For SN1006, we use data from FERMI \citep{2010ApJS..188..405A}
  and H.E.S.S.\ \citep{2010AA...516A..62A} (sum of the two regions). For Vela
  Jr., we adopt gamma-ray data from FERMI \citep{2011ApJ...740L..51T} and
  H.E.S.S. \citep{2018AA...612A...7H}.  For RX J1713, the gamma-ray data are
  taken from FERMI \citep{2011ApJ...734...28A} and
  H.E.S.S. \citep{2018A&A...612A...6H}.}
\label{Fig: 3_spectra}
\end{figure*}

Recent observations of the ambient density around Vela Jr.\ and RX J1713 suggest
the presence of clumps and thus a hadronic origin of the GeV-TeV gamma-ray
emission \citep{2003PASJ...55L..61F,2012ApJ...746...82F, 2012MNRAS.422.2230M,
  2018ApJ...866...76M}. On the other hand, the low density of the dilute phase
of ISM and X-ray measurements for SN1006 suggest mixed leptonic-hadronic
models to explain both the GeV and the TeV flux in a unified picture as shown by
recent simulations \citep{Winner}: while the GeV gamma-ray regime has a
significant leptonic contribution, in this model the TeV range is dominated by
hadronic gamma rays \citep{2010AA...516A..62A}.

We can reproduce the observed VHE gamma-ray spectra of all three SNRs for our
adopted parameters. To demonstrate this, we compare the observational data from
FERMI and H.E.S.S.\ to a one-zone model in which the CR proton spectrum is
described by a power law with exponential cutoff of the form:
\begin{equation}
f^{\rmn{1D}}(p) = \dfrac{\de^2 N}{\de p \de V} \propto p^{-\alpha} \exp\left[ -\left( \dfrac{p}{p_{\mathrm{cut}}} \right)^{\beta} \right]
\end{equation}
where $f^{\rmn{1D}}(p)=4\pi\,p^2f^{\rmn{3D}}(p)$, $\alpha$ is the spectral
index, $p_{\mathrm{cut}}$ is the cutoff momentum and $\beta$ describes the
sharpness of the cutoff; with values reported in Table~\ref{table:Table 1}. 

The resulting spectra of our hadronic models are shown in Fig.~\ref{Fig:
  3_spectra} and match the observed spectra. We further notice that in SN1006
the gamma-ray spectrum extends to higher energies, arguing for a larger maximum
CR proton energy in comparison to the turbulent cases of Vela Jr. and RX
J1713. This difference depends on a range of different factors such the
progenitor (SNIa for SN1006 and core-collapse for Vela Jr. and RX J1713), the
ISM density, the local magnetic field amplification in the upstream and the time
the particles spent in favourable conditions (e.g., quasi-parallel shock
geometries) at the shock. However, the limited angular resolution of
H.E.S.S. precludes a more in-depth analysis of the acceleration mechanism
leaving this task to the next generation of ground-based arrays such the Cherenkov
Telescope.

\section{Conclusions}
\label{sec: conclusions}
In this paper we use MHD simulations with CR physics to explore the effect of
density inhomogeneities on the $\TeV$ gamma-ray morphology from SNRs during
their Sedov-Taylor stage. Our setup allow us to explore several combinations of
homogeneous and turbulent magnetic fields and ambient density distributions. We
find that a single physical model, namely obliquity-dependent shock acceleration
of CRs, is capable of explaining the apparently disparate TeV gamma-ray
morphologies of well-known shell-type SNRs. In this hadronic emission scenario,
gamma-ray bright regions result from quasi-parallel shocks which are known to
efficiently accelerate CR protons, and gamma-ray dark regions point to
quasi-perpendicular shock configurations.

The main characteristics of the emission of SN1006 (a type Ia SN) can be
explained by a homogeneous magnetic field superposed on a density gradient that
explains the different integrated gamma-ray flux of both polar caps. By
contrast, the irregular gamma-ray morphologies of the core collapse SNRs Vela
Jr.\ and RX J1713 is owing to a turbulent magnetic field with
$\vect{B}_0\approx\mathbf{0}$ that is supplemented with a population of
multiphase dense molecular clumps that are characteristic for star formation
regions (and complemented with a weak density gradient in the case of RX
J1713). Adapting a straw man's model of isotropic CR acceleration (that does not
depend on magnetic pre-shock orientation) we conclude that this model is not
able to reproduce the sharp bi-lobed morphology observed for SN1006 even the
presence of moderately strong density fluctuations. Moreover, the simulated
azimuthal profiles of this isotropic acceleration model with strong density
variations cannot reproduce the observed rapid variations of the gamma-ray
emissivity of Vela Jr.\ and RX J1713 without significantly corrugating the shock
surface, which is then ruled out by the spherical morphologies of these SNRs.

Our main findings are summarised here:
\begin{itemize}
\item Moderate density fluctuations can be responsible for a local modulation of
  the gamma-ray emissivity irrespective of the magnetic morphology both in a
  constructive and destructive way. We show that density fluctuations on a scale
  comparable to the size of the remnant and with an amplitude that is stronger
  than $75$ per cent with respect to the mean ISM density causes a corrugated
  shock front that generates strong local variations in the shock acceleration
  efficiency and eventually a very asymmetrical appearance of the gamma-ray SNR.
\item For SN1006, using the relative brightness of the NE and SW lobes, we
  constrain the intensity of the density gradient to be no more than
  $0.0035~\cm^{-3} / \pc$ and directed from SW to NE. We predict that local
  density fluctuations with $\lambda_\rho \simeq 4~\pc$ and $\delta \rho/ \rho =
  0.5$ are secondary in shaping the morphology of the remnant if compared to a
  moderate level of turbulence of the local magnetic field with a coherence
  scale at $1/3$ of the size of the remnant.  However, we note that the presence
  of density fluctuations can explain the strong asymmetry of the distribution
  of Fe and other heavy elements for this remnant.
\item The strong noise level surrounding the SN1006 SNR does not allow more
  precise constraints on the properties of the surrounding ISM.  We conclude
  that for such level of noise a model with negligible density fluctuations
  better represents the morphology of SN1006.  Performing a 3D rotation of the
  SNR around the axis perpendicular to the orientation of $\vect{B}_0$ into the
  line of sight, the emerging radial emission profiles constrain the inclination
  of the magnetic field to be $\lesssim 10^\circ$.  Our azimuthal emission
  profiles for different magnetic inclinations are rather robust and show an
  excellent agreement with the observational profile except for the NE lobe,
  which is somewhat broader in our simulations. While this could signal a
  different functional form of the obliquity dependence of CR acceleration, this
  conclusion is unfortunately degenerate with density fluctuations that could
  also cause a sharper gamma-ray peak.
\item Our generated gamma-ray mock maps with obliquity-dependent acceleration
  are capable of reproducing most of the properties observed for Vela Jr. and RX
  J1713 such the length of the shell filaments, the internal patchy emission,
  the large-scale gamma-ray bright rims and moderate (small-scale) corrugations
  at the shock front. The fainter emission at the centre of the remnants
  expanding in a turbulent magnetic field is compensated by the addition of
  molecular clumps and superposing noise to the images. By contrast an isotropic
  CR acceleration scenario fails to reproduce the azimuthal profiles.
\item In addition, by considering an isotropic CR acceleration scenario for Vela
  Jr. and RX J1713 with a varying level of density fluctuations and coherence
  scales, we exclude strong density fluctuations with a coherence scale
  comparable to the size of the remnant are responsible for the observed
  emission morphology. This is because sufficiently strong density fluctuations
  (that would be needed to explain the significant gamma-ray brightness
  fluctuations) cause a heavily corrugated shock surface which is in direct
  conflict with the almost spherical shape of SNRs RX J1713 and Vela Jr.  The
  comparison between the radial dispersions of our simulated mock maps and the
  excess maps of the SNRs RX J1713 and Vela Jr.\ enables us to limit the density
  fluctuations to $\delta\rho/\rho_0\lesssim75$ per cent of the average ISM
  density.

\end{itemize}

For the first time, our models are able to match morphological and spectral
properties of all known shell-type TeV gamma-ray SNRs. Remarkable improvements
in the angular resolution in future surveys (beyond what is achievable by CTA)
are needed to resolve the TeV emission from individual molecular clumps which
would yield a deeper insight in the structure of the circumstellar ISM.

\section{Acknowledgements}
We would like to thank Ralf Klessen and Volker Springel for the fruitful
comments and suggestions to this work.  We thank our anonymous referee for a
  constructive report that helped to improve the paper.  We acknowledge support
by the European Research Council under ERC-CoG grant CRAGSMAN-646955.

\section{Data availability}
The data underlying this article will be shared on reasonable request to the corresponding author.

\bibliographystyle{mnras}
\bibliography{ms}

\appendix

\newpage

\section{Noise threshold for gamma-ray excess maps}
\label{sec:app}

\begin{figure*}
\includegraphics[width=\textwidth]{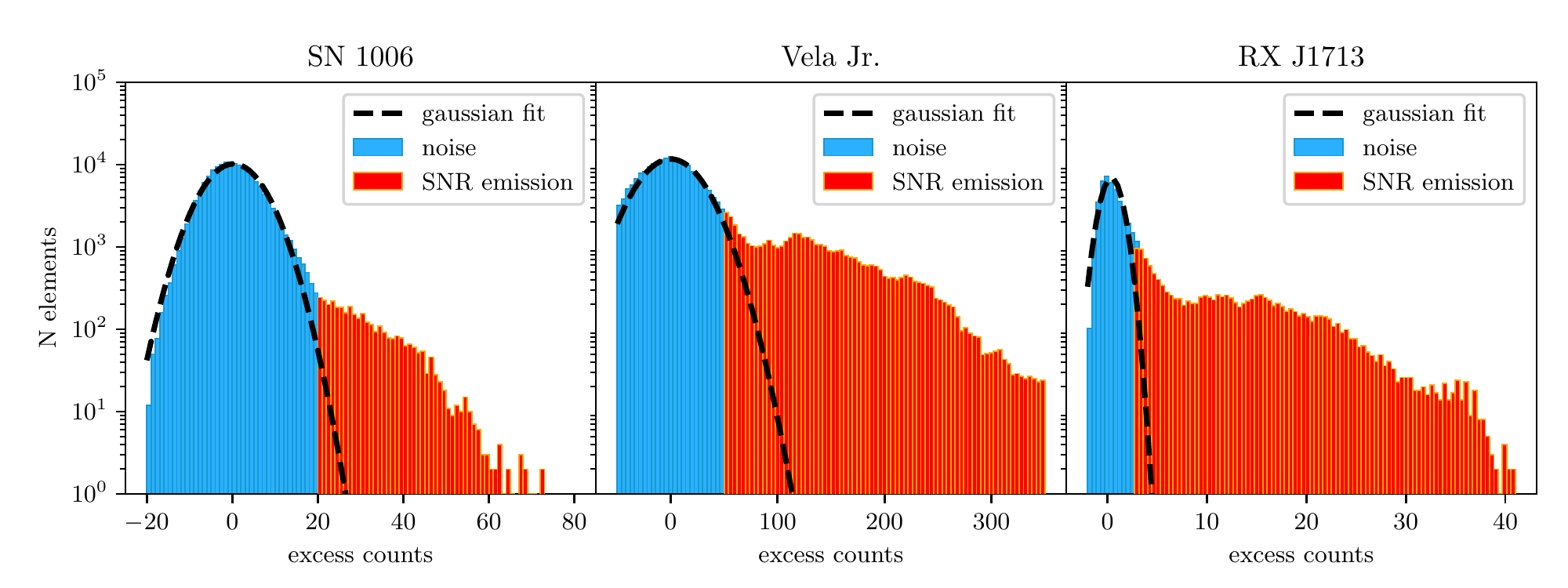} 
\caption{H.E.S.S.\ gamma-ray excess count distributions of the full gamma-ray
    brightness maps of SN 1006 \citep{2010AA...516A..62A}, Vela Jr
    \citep{2018AA...612A...7H}, and RX J1713 \citep{2018A&A...612A...6H}. The
    blue parts of the histograms represent the counts associated with noise
    while the red parts denote excess counts associated with the SNR signal.
    We fit a normal distribution (black dashed line) with zero mean to show
    that the significance distribution of the field-of-view is compatible with
    statistical noise fluctuations.}
\label{Fig: 3_thesholds}
\end{figure*}

The gamma-ray excess maps of SN 1006, Vela Jr. and RX-J1713 show detailed
  signal morphologies in addition to instrumental and gamma-ray reconstruction
  noise.  We assume that the noise is normally distributed and define the noise
  threshold to be equal to the absolute value of the most negative value. To
  justify this approach we report the distribution of excess counts of
  H.E.S.S.\ observations of the three SNRs analysed here in Fig.~\ref{Fig:
    3_thesholds}. This shows that the excess counts at the low end are dominated
  by Gaussian noise with zero mean. We exclude these counts and isolate the
  excess counts of the SNR signal. We set $c_\mathrm{min} = -20$ for SN 1006, $c_\mathrm{min} = -50$ for Vela Jr., and
  $c_\mathrm{min} = -2$ for RX J1713. We note that in our synthetic modeling, the
  generated noise is also normally distributed with zero mean. 

\end{document}